\documentclass{tlp}

\usepackage{aopmath}
\usepackage{latexsym}
\usepackage{amssymb}

\newtheorem{definition}{Definition} 
\newtheorem{example}{Example} 

\makeatletter
\newenvironment{prog}{\vspace{0.7ex}\par
\setlength{\parindent}{0.7cm}
\obeylines\@vobeyspaces\tt}{\vspace{0.7ex}\noindent
}
\makeatother
\newcommand{\startprog}{\begin{prog}}
\newcommand{\stopprog}{\end{prog}\noindent}

\newcommand{\pos}{{\cP}os}
\newcommand{\fpos}{{\cN}{\cV}{\cP}os}
\newcommand{\var}{{\cV}ar}

\newcommand{\sdom}{{\cD}om}
\newcommand{\nat}{\mbox{$I\!\!N$}}

\newcommand{\Cc}{{\cal{C}}} 
\newcommand{\Fc}{{\cal{F}}} 
\newcommand{\Var}{{\cal V}ar} 
\newcommand{\Rc}{{\cal{R}}} 
\newcommand{\Tc}{{\cal{T}}} 
\newcommand{\Xc}{{\cal{X}}} 
\renewcommand{\emptyset}{\varnothing}
\renewcommand{\phi}{\varphi}
\newcommand{\sleq}{\leqslant}
\newcommand{\equ}{\approx}
\newcommand{\dt}{{\cal P}} 
\newcommand{\nns}{\lambda} 
\newcommand{\lns}{\lambda_{lazy}} 
\newcommand{\toppos}{\mbox{\footnotesize$\Lambda$}} 
\newcommand{\ol}[1]{\overline{#1}}  
\newcommand{\pr}[1]{\mbox{$\tt #1$}}   

\newcommand{\traces}{{\backslash\!\!\backslash}}

\newcommand{\inr}{{\to_{>\Lambda}}}
\newcommand{\inrs}{{\to^*_{>\Lambda}}}
\newcommand{\exr}{{\to_{\Lambda}}}

\def\res{\mathrel{\vert\grave{ }}}
\let\l=\langle
\let\r=\rangle
\def \tuple#1{\langle #1 \rangle}

\def\defemb#1#2{\expandafter\def\csname #1\endcsname
                              {\relax\ifmmode #2\else\hbox{$#2$}\fi}}
\defemb{cA}{{\cal A}}
\defemb{cB}{{\cal B}}
\defemb{cC}{{\cal C}}
\defemb{cD}{{\cal D}}
\defemb{cE}{{\cal E}}
\defemb{cF}{{\cal F}}
\defemb{cG}{{\cal G}}
\defemb{cH}{{\cal H}}
\defemb{cI}{{\cal I}}
\defemb{cJ}{{\cal J}}
\defemb{cL}{{\cal L}}
\defemb{cM}{{\cal M}}
\defemb{cN}{{\cal N}}
\defemb{cO}{{\cal O}}
\defemb{cP}{{\cal P}}
\defemb{cR}{{\cal R}}
\defemb{cS}{{\cal S}}
\defemb{cT}{{\cal T}}
\defemb{cU}{{\cal U}}
\defemb{cV}{{\cal V}}
\defemb{cX}{{\cal X}}
\defemb{cZ}{{\cal Z}}

\submitted{1 February 2000}
\revised{25 November 2002, 3 February 2004 }
\accepted{2 March 2004}

\begin{document}

\title[Specialization of Functional Logic Programs Based on Needed
Narrowing]
{Specialization of Functional Logic Programs Based on Needed
Narrowing\thanks{%
  A preliminary short version of this paper appeared in the
  Proceedings of the International Conference on Functional
  Programming (ICFP'99), pp.~273--283, Paris, 1999.
  This paper has been accepted for publication in the Journal of
  Theory and Practice of Logic Programming. In contrast to the
  journal version, this paper contains the detailed proofs of the
  results presented in this paper.
  This work has been
  partially supported by CICYT TIC2001-2705-C03-01, by MCYT under
  grant HA2001-0059, and by the German Research Council (DFG) under
  grant Ha 2457/1-2.}
}

\author[M. Alpuente et al.]
{
MAR\'IA ALPUENTE, SALVADOR LUCAS, GERM\'AN VIDAL\\
DSIC, Technical University of Valencia\\ 
Camino de Vera s/n, E-46020 Valencia, Spain\\ 
\email{$\{$alpuente,slucas,gvidal$\}$@dsic.upv.es}
\and
MICHAEL HANUS \\
Institut f\"ur Informatik \\
CAU Kiel, Olshausenstr.~40, D-24098 Kiel, Germany\\
\email{mh@informatik.uni-kiel.de}
}

\maketitle

\begin{abstract}
Many functional logic languages are based on narrowing,
a unification-based goal-solving mechanism
which subsumes the reduction mechanism of functional languages and
the resolution principle of logic languages. Needed narrowing is
an optimal evaluation strategy which constitutes the basis of modern 
(narrowing-based) lazy functional logic languages. 
In this work, we present 
the fundamentals of partial evaluation in such languages. We provide 
correctness results for partial evaluation based on needed 
narrowing and show that the nice properties of this strategy are 
essential for the specialization process. In particular, the 
structure of the original program is preserved by partial evaluation
and, thus, the same evaluation strategy can be applied for the 
execution of 
specialized programs. This is in contrast to other partial evaluation
schemes for lazy functional logic programs which may change the 
program structure in a negative way.
Recent proposals for the partial evaluation of declarative 
multi-paradigm programs use (some form of) needed narrowing to perform 
computations at partial evaluation time. Therefore, our results 
constitute the basis for the correctness of such partial evaluators.
\end{abstract}


\section{Introduction}
\label{sec-intro}

Functional logic languages combine the operational
principles of the most important declarative programming
paradigms, namely functional and logic programming.
Efficient demand-driven functional computations  are
amalgamated with the flexible use of logical variables
providing for function inversion and search for solutions.
The operational semantics of such languages
is usually based on narrowing,
a generalization of term rewriting which
combines reduction and variable instantiation.
A \emph{narrowing step} instantiates variables of an expression
and applies a reduction step to a \emph{redex} (\emph{red}ucible 
\emph{ex}pression) of the instantiated expression.
The instantiation of variables is usually computed by unifying
a subterm of the entire expression with the left-hand side of some
rule.

\begin{example}
\label{example-leq}
Consider the following rules which define the less-or-equal
predicate ``$\sleq$'' on natural numbers which are represented by terms
built from data constructors \pr{0} and \pr{s} (note that variable names 
always start with an uppercase letter):
\[
\begin{array}{r@{~\sleq~}l@{~~\rightarrow~~}l}
\tt 0    & \tt N & \tt true \\
\tt s(M) & \tt 0 & \tt false \\
\tt s(M) & \tt s(N) & \tt M \sleq  N
\end{array}
\]
The goal $\tt s(X) \sleq Y$ can be solved (i.e., reduced to \pr{true}) by
instantiating \pr{Y} to \pr{s(Y1)} to apply the third rule followed by the
instantiation of \pr{X} to \pr{0} to apply the first rule:
\[\tt
s(X) \sleq Y  ~~\leadsto_{\{Y \mapsto s(Y1)\}}~~
X \sleq Y1 ~~\leadsto_{\{X \mapsto 0\}}~~
true
\]
\end{example}
Narrowing provides completeness in the sense of logic programming
(computation of all solutions) as well as functional programming
(computation of values). Since simple narrowing can have a
huge search space, great effort has been made to develop
sophisticated narrowing strategies without losing completeness;
see \cite{Han94JLP} for a survey.
To avoid unnecessary computations and to provide computations
with infinite data structures as well as a demand-driven
generation of the search space, most recent work has advocated
\emph{lazy} narrowing strategies, e.g., 
\cite{AEH00,GLMP91,LLR93,MR92}.
Many lazy evaluation strategies are based on
the notions of \emph{demanded} or \emph{needed} computations. 
The following example informally explains the difference between 
these two notions:

\begin{example}\label{ExDemandedVersusNeeded}
Consider the rules for ``$\sleq$'' in Example~\ref{example-leq}
together with the following rules defining the addition on natural numbers:
\[
\begin{array}{r@{~~\rightarrow~~}l}
\tt 0    + N & \tt N \\
\tt s(M) + N & \tt s(M+N)
\end{array}
\]
The initial term is $\tt X \sleq X+X$. The evaluation of subterm
\pr{X+X} is \emph{demanded} by the second and third rules for
``$\sleq$'', since these rules cannot be applied to $\tt X \sleq X+X$ 
until the subterm \pr{X+X} is reduced to a term rooted by a data 
constructor symbol. 
However, evaluating this subterm is not \emph{needed}
since, if we instantiate \pr{X} to \pr{0}, we directly obtain
\pr{true} by using the first rule for ``$\sleq$\rlap{.}''

On the other hand, if the initial term is \pr{X+(0+0)}, the evaluation
of \pr{0+0} is \emph{needed} to compute its value whereas it is not
\emph{demanded} by any rule for ``\pr{+}\rlap{.}''
\end{example}
\emph{Needed narrowing} \cite{AEH00} is based on the idea
of evaluating only subterms which are \emph{needed} in order to compute
a result. For instance, in a term like $t_1 \sleq t_2$,
it is always necessary to evaluate $t_1$ (to some \emph{head normal 
form}, i.e., either a variable or a constructor-rooted term)
since all three rules in Example~\ref{example-leq} have left-hand
sides whose first argument is not a variable.
On the other hand, the evaluation of $t_2$ is only
needed if $t_1$ is of the form \pr{s(\cdots)}.
Thus, if $t_1$ is a free variable, needed narrowing instantiates
it to a constructor, here \pr{0} or \pr{s(\cdots)}.
Then, depending on this instantiation,
either the first rule is applied or the second argument $t_2$ is evaluated.
Needed narrowing is currently the best
narrowing strategy for first-order functional logic programs due to
its optimality properties w.r.t.\ the length of derivations and the
number of computed solutions \cite{AEH00}.
Informally speaking, needed narrowing derivations are the shortest possible
narrowing derivations if common subterms are shared (as it is usually
done in implementations of functional languages),
and the set of all solutions computed by needed narrowing
is minimal since needed narrowing computes only independent solutions
(see also Theorem~\ref{theo-needed-properties} below).
Furthermore, it can be efficiently implemented by pattern matching
and unification \cite{Han95b,LLR93}. For instance, the operational semantics 
of the declarative multi-paradigm language Curry \cite{CurryTR} is based 
on needed narrowing. Needed narrowing has also been
extended to higher-order functions and $\lambda$-terms as data
structures and proved optimal w.r.t.\ the independence of computed
solutions \cite{HP99}. 

Partial evaluation (PE) is a semantics-preserving performance
optimization technique for computer programs which consists of the
specialization of the program w.r.t.\ parts of its input.
PE has been widely applied in the fields of
term rewriting systems \cite{Bel95,Bon88,DR93,LG97b},
functional programming
\cite{CD93,JGS93},  and logic programming \cite{Gal93,LS91,dSGJLMS99}.
Although the objectives
are similar, the general methods are often different due to the
distinct underlying models and the different perspectives
\cite{AFV98}.
This separation has the negative consequence of
duplicated work since developments are not shared and many similarities
are overlooked.
A unified treatment can bring the different
methodologies closer and lays the ground for new insights
in all three fields \cite{AFV98,AFV98b,GS94,PP96,SGJ93}.

In order to perform reductions at specialization time,
\emph{online} partial evaluators normally include an interpreter
\cite{CD93}.
This implies that the power of the transformation is highly influenced
by the properties of the evaluation strategy in the underlying
interpreter.
Narrowing-driven PE \cite{AFV98,AV02} is the first generic algorithm for
the specialization of functional logic programs. The method is
parametric w.r.t.\ the narrowing strategy which is used
for the automatic construction of the search trees.
The method is formalized within the theoretical framework established 
by \citeN{LS91} for the PE of logic programs
(also known as \emph{partial deduction}), although
a number of concepts have been generalized to
deal with the functional component of the language (e.g., nested function
calls in expressions, different evaluation strategies, etc).
This approach has better opportunities for optimization thanks to the
functional dimension (e.g., by the inclusion of deterministic
evaluation steps). Also, since unification is embedded into
narrowing, it is able to automatically propagate syntactic information on the
partial input (term structure) and not only constant values, similar
to partial deduction.
Using the terminology of
\citeN{GS96}, narrowing-driven PE is able to produce both
\emph{polyvariant} and \emph{polygenetic} specializations,
i.e., it can produce different specializations for the same
function definition
and can also combine distinct original function definitions
into a comprehensive specialized function.
This means that narrowing-driven PE has the same potential for
specialization as \emph{positive supercompilation}
of functional programs \cite{SGJ93} and \emph{conjunctive partial
deduction} of logic programs \cite{dSGJLMS99}; more detailed comparisons 
can be found in \cite{AFV98,AFV98b,AV02}.

The main contribution of this work is the proof of
the basic computational properties of PE based on needed narrowing.
The most recent approaches for the PE of multi-paradigm
functional logic languages \cite{AAHV99,AHV02,AHV02b}
use (a form of) needed narrowing to perform computations at PE time
(see also Section~\ref{extensions}).
Therefore, our results 
constitute the basis for the correctness of such partial evaluators.
To be more precise, we provide the following results for PE based on
needed narrowing:
\begin{itemize}
\item We prove the strong correctness of the PE scheme:
the answers computed by needed narrowing in the original and the
partially evaluated programs coincide.

\item We establish the relation between PE based on needed narrowing 
and PE based on a different lazy evaluation mechanism---which 
is the basis of previous partial evaluators \cite{AFJV97}. We 
formally prove the superiority of needed narrowing to perform 
partial computations. In particular, we prove that the 
structure of the original program is preserved by PE based on needed 
narrowing and, thus, the same optimal evaluation strategy can be applied 
for the execution of specialized programs. This is in contrast to 
previous PE schemes \cite{AFJV97} for lazy functional logic programs 
which may change the program structure in a negative way.

\item We show that specialized programs preserve deterministic
evaluations,
i.e., if the source program can evaluate a  goal without any choice,
then the partially evaluated program does just the same.
This is important from an implementation point of view and it is not obtained
by PE based on other operational models, like lazy narrowing.
\end{itemize}
Providing experimental evidence of the practical advantages of using 
needed narrowing to perform PE is outside the scope of this paper. 
We refer, e.g., to \cite{AHV02} where this topic has been extensively
addressed for a practical partial evaluator based on the
foundations presented in this paper.

The structure of the paper is as follows. After some basic
definitions in the next section, we recall in Section~\ref{sec-nn}
the formal definition of inductively sequential programs and
needed narrowing. Section~\ref{sec-ln} recalls the lazy narrowing
strategy and relates it to needed narrowing. The definition of
partial evaluation based on needed narrowing is provided in
Section~\ref{sec-nnpe} together with results about the structure
of specialized programs and the (strong) correctness of the
transformation. Section~\ref{extensions} outlines several recent 
extensions of PE based on needed narrowing. Finally, 
Section~\ref{sec-concl} concludes. Proofs of selected results can be 
found in an appendix.

\section{Preliminaries}

Term rewriting systems (TRSs) provide an adequate computational model for
functional languages which allow the definition of functions by means 
of patterns (e.g., Haskell, Hope or Miranda).
Within this framework,
the class of \emph{inductively sequential} programs, which we consider
in this paper, has been defined,
studied, and used for the implementation of programming languages which
provide for optimal computations both in functional and functional
logic programming \cite{Ant92,AEH00,Han97b,HanLucMid98,LLR93}. Inductively
sequential programs can be thought of as constructor-based TRSs with
discriminating left-hand sides, i.e., typical functional programs
where at most one rule is used to reduce a particular subterm
(without variables).
Thus, in the remainder of the paper
we follow the standard framework of term rewriting \cite{DJ90}
for developing our results.

We consider a (\emph{many-sorted}) \emph{signature}
$\Sigma$ partitioned into a set
$\Cc$ of \emph{constructors} and a set $\Fc$ of (defined)
\emph{functions} or \emph{operations}. We write
$c/n \in \Cc$ and $f/n \in \Fc$ for $n$-ary constructor and operation
symbols, respectively. There is at least one sort $Bool$ containing
the Boolean constructors $true$ and $false$. Given a set of
variables $\Xc$, the set of
\emph{terms} and \emph{constructor terms}
are denoted by $\Tc(\Cc \cup \Fc,\Xc)$ and $\Tc(\Cc,\Xc)$, respectively.
The set of variables occurring in a term $t$ is denoted by $\Var(t)$.
A term $t$ is \emph{ground} if $\Var(t) = \emptyset$.
A term is \emph{linear} if it does not contain multiple occurrences
of one variable.
We write $\ol{o_n}$ for the \emph{sequence} of objects $o_1,\ldots,o_n$.

A \emph{pattern} is a term of the form $f(\ol{d_n})$
where $f/n \in \Fc$ and $d_1,\ldots,d_n \in \Tc(\Cc,\Xc)$.
A term is \emph{operation-rooted} if it has an operation symbol
at the root.
$root(t)$ denotes the symbol at the root of the term $t$.
A \emph{position} $p$ in a term $t$ is represented by a sequence of
natural numbers ($\toppos$ denotes the empty sequence, i.e., the
root position).
They are used to address the nodes of a term viewed as a tree (Dewey
notation). For instance, if $t=f(t_1,\ldots,t_n)$, positions
$1,\ldots,n$ refer to arguments $t_1,\ldots,t_n$ respectively; thus,
given a position $p_i$ of a subterm of $t_i$, position $i.p_i$ denotes the
corresponding subterm of $t$.
Positions are ordered by the \emph{prefix} ordering: $u \leq v$, if there
exists $w$ such that $u.w =v$.
Given a term $t$, $\pos(t)$ and $\fpos(t)$ denote the set of
positions and the set of non-variable positions of $t$, respectively.
$t|_p$ denotes the \emph{subterm} of $t$ at position
$p$, and $t[s]_p$ denotes the result of
\emph{replacing the subterm} $t|_p$ by the term $s$
(see \cite{DJ90} for details).

We denote by $\{x_1 \mapsto t_1,\ldots, x_n \mapsto t_n\}$ the
\emph{substitution} $\sigma$ with
$\sigma(x_i) = t_i$  for $i=1,\ldots,n$ (with $x_{i}\neq
x_{j}$ if  $i\neq
j$), and $\sigma(x) = x$ for all other variables $x$.
The set $\sdom(\sigma) =
\{x \in \cX \mid \sigma(x) \neq x\}$ is called the \emph{domain} of
$\sigma$.
A substitution $\sigma$ is \emph{constructor} (\emph{ground constructor}),
if $\sigma(x)$ is constructor (ground constructor)
for all $x \in \sdom(\sigma)$.
The identity substitution is denoted by $id$.
Substitutions are extended to morphisms on terms by
$\sigma(f(\ol{t_n})) = f(\ol{\sigma(t_n)})$
for every term $f(\ol{t_n})$.
Given a substitution $\theta$ and a set of variables
$V \subseteq \Xc$, we denote by $\theta {}_{\res V}$ the substitution
obtained from $\theta$ by restricting its domain to $V$. We write
$(\theta = \sigma)~[V]$ if $\theta {}_{\res V} = \sigma {}_{\res V}$,
and $(\theta \leq \sigma)~[V]$ denotes the existence of a
substitution $\gamma$ such that $(\gamma\circ\theta = \sigma)~[V]$.

Term $t'$ is an \emph{instance} of $t$ if there is a
substitution $\sigma$ with $t' = \sigma(t)$. This implies a
(relative generality) \emph{subsumption ordering} on terms which
is defined by $t \leq t'$ iff $t'$ is an instance of $t$. A
\emph{unifier} of two terms $s$ and $t$ is a substitution $\sigma$
with $\sigma(s) = \sigma(t)$. The unifier $\sigma$ is \emph{most
general} if $(\sigma \leq \sigma')~[\Xc]$ for each other unifier
$\sigma'$. 

A rewrite rule is an ordered pair $(l,r)$, written $l\to
r$,  with $l,r\in\Tc(\Cc \cup \Fc,\Xc)$, $l\not\in \Xc$ and
$\Var(r)\subseteq \Var(l)$.
A set of rewrite rules
is called a \emph{term rewriting system}
(TRS). The terms
$l$ and $r$ are called the \emph{left-hand side} (\emph{lhs}) and the
\emph{right-hand side} (\emph{rhs}) of the rule,
respectively. A
TRS $\cR$ is \emph{left-linear} if $l$ is linear for all $l\to r\in\cR$.
A TRS is constructor-based (CB) if each lhs $l$ is a pattern.
Two (possibly renamed) rules $l\to r$  and $l'\to r'$ \emph{overlap},
if there is a non-variable
position $p\in\fpos(l)$ and a most general unifier $\sigma$ such that
$\sigma(l|_p)=\sigma(l')$.
A left-linear TRS without overlapping rules is
called \emph{orthogonal}.
In the remainder of this paper, a \emph{functional logic program} is a
finite left-linear CB-TRS.
Conditions in program rules are treated by using
the predefined functions \pr{and}, \pr{if\_then\_else}, \pr{case\_of}
which are reduced by standard defining rules \cite{MR92}.

A \emph{rewrite step} is an application of a rewrite rule to a term, i.e.,
$t \to_{p,R} s$ if there is a position $p$ in $t$,
a rewrite rule $R$ of the form $l \to r$ and a substitution $\sigma$
with $t|_p = \sigma(l)$ and $s = t[\sigma(r)]_p$
($p$ and $R$ will often be omitted in the notation of a
rewrite step).
The instantiated lhs $\sigma(l)$ is called a \emph{redex}.
${\cal P}os_{\cal R}(t)$ denotes the set of \emph{redex
positions} of the term $t$ in the TRS $\Rc$.
$\rightarrow^{+}$ ($\rightarrow^{\ast}$) denotes the transitive
(reflexive and transitive) closure
of $\rightarrow$. If $t\to^* s$, we say that $t$
is rewritten to $s$.
A term $t$ is root-stable (often called a \emph{head-normal form})
if it cannot be rewritten
to a redex. 
A \emph{constructor root-stable} term is either a variable or a
\emph{constructor-rooted}
term, i.e., a term rooted by a constructor symbol. A term $t$ is called
\emph{irreducible} or in \emph{normal form} if there is no term $s$
with $t \to s$.

In order to evaluate terms containing variables, narrowing
non-deterministically
instantiates its variables such that a rewrite step is 
possible---usually by computing most general unifiers between a subterm and
some lhs \cite{Han94JLP}, but this requirement is relaxed
in needed narrowing steps in order to obtain an optimal evaluation
strategy \cite{AEH00}. 
Formally,
$t \leadsto_{p,R,\sigma} t'$ is a \emph{narrowing step}
if $p$ is a non-variable position in $t$ and $\sigma(t) \to_{p,R} t'$.
We denote by $t_0 \leadsto^\ast_\sigma t_n$ a sequence of narrowing
steps $t_0 \leadsto_{\sigma_1} \ldots \leadsto_{\sigma_n} t_n$
with $\sigma = \sigma_n \circ \cdots \circ \sigma_1$
(if $n=0$ then $\sigma = id$).
Since we are interested in computing \emph{values} (constructor terms)
as well as \emph{answers} (substitutions), 
we say that the narrowing derivation $t \leadsto^\ast_\sigma c$
\emph{computes the result $c$ with answer $\sigma$}
if $c$ is a constructor term.
The evaluation to ground constructor terms 
is the most common semantics of functional (logic) languages.
In lazy functional (logic) languages, the equality predicate $\equ$
used in some examples is defined
as the \emph{strict equality} on terms (note that we do not require
terminating rewrite systems and, thus, reflexivity is not desired),
i.e., the equation $t_1 \equ t_2$ is satisfied if and only if
$t_1$ and $t_2$ are reducible to the same ground constructor term.
Furthermore, a substitution $\sigma$ is a \emph{solution} for an
equation $t_1 \equ t_2$ if $\sigma(t_1) \equ \sigma(t_2)$ is
satisfied. The strict equality can be defined as a binary Boolean
function by the following set of orthogonal rewrite rules:
\[ \hspace{-6ex}
\begin{array}{rcll}
\tt c \approx c & \rightarrow & \tt true &
~~ \tt c/0 \in \cC \\
\tt c(X_{1},\ldots,X_{n}) \approx c(Y_{1},\ldots,Y_{n}) & \rightarrow &
\tt (X_{1} \approx Y_{1}) \wedge \ldots \wedge (X_{n} \approx Y_{n}) &
~~ \tt c/n \in \cC, n>0 \\
\tt true \wedge X & \rightarrow & \tt X  & \\
\end{array} 
\]
Thus, we do not treat strict equality  in any
special way and it is sufficient
to consider it as a Boolean function.
We say that $\sigma$ is a \emph{computed answer substitution}
for an equation $e$ if there is a narrowing derivation
$e \leadsto^{\ast}_{\sigma} \tt true$.
More details about strict equality can be found in
\cite{AEH00,GLMP91,MR92}.

As in logic programming, narrowing derivations can be represented by a
(possibly infinite) finitely branching \emph{tree}.  Formally, given a
program $\cR$ and an operation-rooted term $t$, a \emph{narrowing
  tree} for $t$ in $\cR$ is a tree satisfying the following
conditions: (a) each node of the tree is a term, (b) the root node is
$t$, and (c) if $s$ is a node of the tree then, for each narrowing
step $s \leadsto_{p,R,\sigma} s'$, the node has a child $s'$ and the
corresponding arc in the tree is labeled with $(p,R,\sigma)$. A \emph{failing
  leaf} contains a term which is not a constructor term and which
cannot be further narrowed.
Following \cite{LS91}, in  this work
we adopt the convention that a derivation can be 
\emph{incomplete} (thus, a branch can be failed, incomplete, successful, 
or infinite).

\section{Needed Narrowing}\label{sec-nn}

Since functional logic languages are intended to extend
(pure) logic languages, completeness of the operational semantics
is an important issue. Similarly to logic programming,
completeness means the ability to compute representatives of
all solutions for one or more equations (this will be formalized in
Theorem~\ref{theo-needed-properties}).
Narrowing, as defined in the previous section,
is complete but highly (don't-know) non-deterministic:
if $t$ is a term, we have to apply at all non-variable subterms
all possible rules with all possible substitutions in order
to compute all solutions. Clearly, this would be too inefficient
for a realistic functional logic language.
Thus, a challenge in the design of functional logic languages is the
definition of a ``good'' narrowing strategy, i.e., a restriction
on the narrowing steps issuing from a given term $t$, without losing
completeness. \cite{Han94JLP}  contains a survey of various
attempts to define reasonable narrowing strategies.

\emph{Needed narrowing} \cite{AEH00} is currently
the best known narrowing strategy due to its optimality properties
(see the discussion in Section~\ref{sec-intro}
and Theorem~\ref{theo-needed-properties}).
Needed narrowing is defined on
\emph{inductively sequential programs}, a class of
CB-TRSs where the left-hand sides do not overlap (in particular,
they are not unifiable). To provide a definition of this class
of programs and the needed narrowing strategy, we introduce
definitional trees \cite{Ant92}.
Here we use the
definition of \cite{Ant97} which is more appropriate for our
purposes.

A \emph{definitional tree} of a finite set $S$ of linear patterns
is a non-empty set $\dt$ of linear patterns partially ordered
by subsumption having the following properties:
\begin{description}
\item[\mbox{\rm\em Root property:}]
$\dt$ has a minimum element (that we denote as $pattern(\dt)$),
also called the \emph{pattern} of the definitional tree.
\item[\mbox{\rm\em Leaves property:}]
The maximal elements of $\dt$, called the \emph{leaves} of the definitional tree, are the
elements of $S$. Non-maximal elements are also called \emph{branch}
nodes.
\item[\mbox{\rm\em Parent property:}]
If $\pi \in \dt$, $\pi \neq pattern(\dt)$,
there exists a unique $\pi' \in \dt$, called the \emph{parent}
of $\pi$ (and $\pi$ is called a \emph{child} of $\pi'$),
such that $\pi'< \pi$ and
there is no other pattern $\pi'' \in \Tc(\Cc \cup \Fc,\Xc)$ with
$\pi' < \pi'' < \pi$.
\item[\mbox{\rm\em Induction property:}]
Given $\pi \in \dt \backslash S$, there is a position
$o$ in $\pi$ with $\pi|_o \in \Xc$ (called the
\emph{inductive position}), and constructors $c_1/k_1,\ldots,c_n/k_n\in\Cc$
with $c_i\neq c_j$ for $i\neq j$, such that, for all
$\pi_1,\ldots,\pi_n$ which have the parent $\pi$,
$\pi_i = \pi[c_i(\ol{x_{k_i}})]_o$ (where $\ol{x_{k_i}}$
are new distinct variables) for all $1 \leq i \leq n$.\footnote{%
There might be more than one potential inductive position
when constructing a definitional tree. In this case one can
select any of them since the results about needed narrowing
do not depend on the selected definitional tree.}
\end{description}
If $\Rc$ is an orthogonal TRS and $f/n$ a defined function, we
call $\dt$ a \emph{definitional tree of $f$} if $pattern(\dt) =
f(\ol{x_n})$ for distinct variables $\ol{x_n}$ and the leaves of
$\dt$ are all (and only) variants of the left-hand sides of the
rules in $\Rc$ defining $f$ (i.e., rules $l\to r$ such that
$root(t)=f$, $f \in {\cal F}$). Due to the orthogonality of $\Rc$,
we can assign a unique rule defining $f$ to each leaf. A defined
function is called \emph{inductively sequential} if it has a
definitional tree. A rewrite system $\Rc$ is called
\emph{inductively sequential} if all its defined functions are
inductively sequential. An inductively sequential TRS can be
viewed as a set of definitional trees, each defining a function
symbol. There can be more than one definitional tree for an
inductively sequential function. In the following, we assume that
there is a fixed definitional tree for each defined function.

\begin{figure}
\begin{center}
\setlength{\unitlength}{.8cm}
$\fbox{
\begin{picture}(12,3.8)
\put(2,1.8){$\tt 0 \sleq  Y \to true$}
\put(6,1.8){$\tt s(X')
\sleq  \fbox{\tt Y}$}
\put(3.5,0.1){$\tt s(X') \sleq  0 \to false$}
\put(7.5,0.1){$\tt s(X') \sleq  s(Y') \to X' \sleq Y'$}
\put(4.5,3.4){$\tt \fbox{\tt X} \sleq  Y$}
\put(7,1.6){\line(-3,-2){1.6}} \put(7,1.5){\line(3,-2){1.6}}
\put(5,3.2){\line(-3,-2){1.5}} \put(5,3.1){\line(3,-2){1.5}}
\end{picture}
}$
\end{center}
\caption{Definitional tree for the function ``$\sleq$''\label{fig-deftree}}
\end{figure}
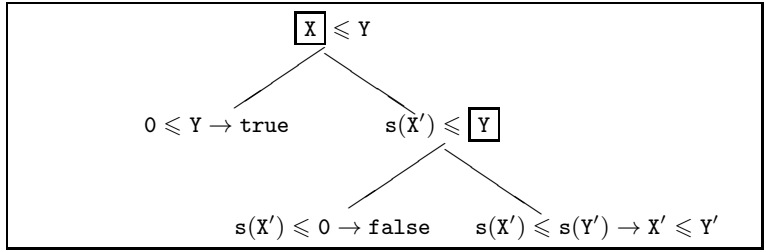

It is often convenient and simplifies understanding to provide a
graphic representation of definitional trees, where each inner
node is marked with a pattern, the inductive position in branch
nodes is surrounded by a box, and the leaves contain the
corresponding rules. For instance, the definitional tree of the
function ``$\sleq$'' in Example~\ref{example-leq} is illustrated
in Figure~\ref{fig-deftree}.

The following auxiliary proposition shows that functions defined by a single
rule are always inductively sequential.

\begin{proposition} \label{prop-dt-linear-pattern}
If $f(\ol{t_n})$ is a linear pattern, then there exists a
definitional tree
for the set $\{ f(\ol{t_n}) \}$ with pattern $f(\ol{x_n})$.
\end{proposition}
\begin{proof}
By induction on the number of constructor symbols occurring in $t$,
where each constructor symbol is introduced in a child of a branch node
and each branch node has only one child.
\end{proof}
For the definition of needed narrowing, we assume that $t$
is an operation-rooted term and $\dt$ is a definitional tree
with $pattern(\dt) = \pi$ such that $\pi \leq t$.
We define a function $\nns$ from terms and definitional trees
to sets of tuples (position, rule, substitution)
as the least set satisfying the following properties.
We consider two cases for $\dt$:\footnote{This description
of a needed narrowing step is slightly different from
\cite{AEH00} but it results in the
same needed narrowing steps.}
\begin{enumerate}
\item If $\pi$ is a leaf, i.e., $\dt = \{\pi\}$, and $\pi \to r$
is a variant of a rewrite rule, then
\[
\nns(t,\dt) = \{(\toppos,\pi \to r,id)\} ~.
\]
\item If $\pi$ is a branch node, consider the inductive position $o$
of $\pi$ and a child $\pi_i = \pi[c_i(\ol{x_n})]_o \in \dt$.
Let $\dt_i = \{ \pi' \in \dt \mid \pi_i \leq \pi' \}$ be
the definitional tree where all patterns are instances of $\pi_i$.
Then we consider the following cases for the subterm $t|_o$:
\[ \hspace{-3ex}
\def\moveon{\phantom{if\ }}
\begin{array}{@{\hspace{-3ex}} r @{\,} l @{}}
\nns(t,\dt) \ni &
\left\{
\begin{array}{@{} l l}
(p,R,\sigma \circ \tau)
 & \mbox{if $t|_o = x \in \Xc$, $\tau = \{x \mapsto c_i(\ol{x_n})\}$,}\\
 & \mbox{\moveon and $(p,R,\sigma) \in \nns(\tau(t),\dt_i)$;} \\[1ex]
(p,R,\sigma \circ id)
 & \mbox{if $t|_o = c_i(\ol{t_n})$ and $(p,R,\sigma) \in \nns(t,\dt_i)$;} \\[1ex]
(o.p,R,\sigma \circ id)
 & \mbox{if $t|_o = f(\ol{t_n})$, $f \in \Fc$, and $(p,R,\sigma) \in
\nns(t|_o,\dt')$} \\
 & \mbox{\moveon where $\dt'$ is a definitional tree for $f$.}
\end{array}
\right.
\end{array}
\]
\end{enumerate}
Informally speaking, needed narrowing applies a rule, if
the definitional tree does not require further pattern matching
(case 1), or checks the subterm corresponding to
the inductive position of the branch node (case 2): if it is a variable,
it is instantiated to the constructor of a child;
if it is already a constructor, we proceed with the corresponding
child (note that we do not actually need substitution $id$ but we
include it to provide a normalized
representation of a needed narrowing step, see below); 
if it is a function, we evaluate it by recursively
applying needed narrowing. Thus, the strategy differs
from typical lazy functional languages only in the 
instantiation of free variables.

Note that, in each recursive step during the computation
of $\nns$, we compose the current substitution with the local substitution of this
step (which can be the identity). Thus, each needed narrowing step
can be represented as $(p,R,\phi_k \circ\cdots\circ \phi_1)$,
where each $\phi_j$ is either the identity or the replacement of a single
variable computed in each recursive step (see the following
proposition). This is also called
the \emph{canonical representation} of a needed narrowing step.
As in proof procedures for logic programming, we assume that the
definitional trees always contain new variables if they are used
in a narrowing step. This implies that all computed
substitutions are idempotent (we will implicitly assume this
property in the following).

To compute needed narrowing steps for an operation-rooted term
$t$, we take the definitional tree $\dt$ for the root of $t$ and
compute $\nns(t,\dt)$. Then, for all $(p,R,\sigma) \in
\nns(t,\dt)$, $t \leadsto_{p,R,\sigma} t'$ is a \emph{needed
narrowing step}. We call this step \emph{deterministic} if
$\nns(t,\dt)$ contains exactly one element.

\begin{example}\label{example-leqadd}
Consider the rules in Example \ref{ExDemandedVersusNeeded}.
Then the function $\nns$ computes the following set for the
initial term $\tt X \sleq X+X$:
\[
\tt \{(\toppos,0 \sleq N \to true, \{X \mapsto 0\}),~~
(2,s(M)+N \to s(M+N), \{X \mapsto s(M)\}) \}
\]
This corresponds to the following narrowing steps:
\[
\begin{array}{lll}
\tt X \sleq X+X & \tt\leadsto_{\{X \mapsto 0\}}    & \tt true \\[1ex]
\tt X \sleq X+X & \tt\leadsto_{\{X \mapsto s(M)\}} & \tt s(M) \sleq s(M+s(M))
\end{array}
\]
\end{example}
In the following we state some interesting properties of
needed narrowing which are useful for our later results.
The first proposition shows that each substitution in a needed
narrowing step instantiates only variables occurring in the
initial term.

\begin{proposition}\label{prop-nn-subst}
If $(p,R,\phi_k \circ\cdots\circ \phi_1) \in \nns(t,\dt)$ is a
needed narrowing step, then, for $i=1,\ldots,k$, $\phi_i = id$ or
$\phi_i = \{x \mapsto c(\ol{x_n})\}$ (where $\ol{x_n}$ are
pairwise different variables) with $x \in
\Var(\phi_{i-1}\circ\cdots\circ\phi_1(t))$.
\end{proposition}
\begin{proof}
By induction on $k$.
\end{proof}
The next lemma shows that for different narrowing steps (computing
different substitutions) there is always a variable which is
instantiated to different constructors:

\begin{lemma} \label{lemma-nn-diff}
Let $t$ be an operation-rooted term, $\dt$ a definitional tree
with $pattern(\dt) \leq t$ and 
$(p,R,\phi_k
\circ\cdots\circ \phi_1), (p',R',\phi'_{k'} \circ\cdots\circ
\phi'_1) \in \nns(t,\dt)$, $k \leq k'$. Then, for all $i \in
\{1,\ldots,k\}$,
\begin{itemize}
\item either $\phi_i\circ\cdots\circ\phi_1 =
\phi'_i\circ\cdots\circ\phi'_1$, or
\item there exists some $j < i$ with
\begin{enumerate}
\item $\phi_j\circ\cdots\circ\phi_1 = \phi'_j\circ\cdots\circ\phi'_1$, and
\item $\phi_{j+1} = \{x \mapsto c(\cdots)\}$ and
      $\phi'_{j+1} = \{x \mapsto c'(\cdots)\}$ with $c \neq c'$.
\end{enumerate}
\end{itemize}
\end{lemma}

\begin{proof}
By induction on $k$ (the number of recursive steps performed by
$\nns$ to compute $(p,R,\phi_k \circ\cdots\circ \phi_1)$):

\begin{description}
\item[$k=1$:]
Then $\dt = \{\pi\}$ and $\nns(t,\dt) = \{(\toppos,R,id)\}$. Thus,
the proposition trivially holds.
\item[$k>1$:]
Then $\pi = pattern(\dt)$ is a branch node and there is an
inductive position $o$ of $\pi$ such that all children of $\pi$
have the form $\pi_i = \pi[c_i(\ol{x_n})]_o \in \dt$. Let $\dt_i =
\{ \pi' \in \dt \mid \pi_i \leq \pi' \}$ be the definitional tree
where all patterns are instances of $\pi_i$, for $i=1,\ldots,n$.
We prove the induction step by a case distinction on the form of
the subterm $t|_o$:
\begin{description}
\item[$t|_o = x \in \Xc$:] Then $\phi_1 = \{x \mapsto c_i(\ol{x_n})\}$
and $(p,R,\phi_k\circ\cdots\circ\phi_2) \in \nns(\phi_1(t),\dt_i)$
for some $i$. If $\phi'_1 = \{ x \mapsto c(\cdots) \}$ with $c
\neq c_i$, then the proposition directly holds. Otherwise, if
$\phi_1 = \phi'_1$, the proposition follows from the induction
hypothesis applied to
$(p,R,\phi_k\circ\cdots\circ\phi_2),(p',R',\phi'_{k'}\circ\cdots\circ\phi'_2)
\in \nns(\phi_1(t),\dt_i)$.
\item[$t|_o = c_i(\ol{t_n})$:]
Then $\phi_1 = id$ and $(p,R,\phi_k\circ\cdots\circ\phi_2) \in
\nns(t,\dt_i)$. Clearly, $\phi'_1 = id$ by definition of $\nns$.
Hence the proposition follows from the induction hypothesis
applied to
$(p,R,\phi_k\circ\cdots\circ\phi_2),(p',R',\phi'_{k'}\circ\cdots\circ\phi'_2)
\in \nns(t,\dt_i)$.
\item[$t|_o = f(\ol{t_n})$:]
Then $\phi_1 = id$ and $(p,R,\phi_k\circ\cdots\circ\phi_2) \in
\nns(t|_o,\dt')$ where $\dt'$ is a definitional tree for $f$. By
definition of $\nns$, $\phi'_1 = id$. Then the proposition follows
from the induction hypothesis applied to
$(p,R,\phi_k\circ\cdots\circ\phi_2),(p',R',\phi'_{k'}\circ\cdots\circ\phi'_2)
\in \nns(t|_o,\dt')$.
\end{description}
\end{description}
\end{proof}
For inductively sequential programs, needed narrowing is sound and complete
w.r.t.\ strict equality when we consider constructor substitutions
as solutions (note that constructor substitutions are sufficient
in practice since a broader class of solutions would contain
unevaluated or undefined expressions for the considered programs).
Moreover, needed narrowing does not compute
redundant solutions. These properties are formalized as follows, 
where we say that two substitutions $\sigma$ and $\sigma'$ are
\emph{independent} (on a set of variables $V\subseteq\Xc$) iff
there is some $x \in V$ such that $\sigma(x)$ and $\sigma'(x)$
are not unifiable.\footnote{Actually, \cite{AEH00} prove a stronger property
(disjointness of solutions) but this is not necessary here.}

\begin{theorem}[\citeNP{AEH00}] \label{theo-needed-properties}
Let $\Rc$ be an inductively sequential program and $e$ an
equation.
\begin{enumerate}
\item (Soundness) If $e \leadsto^\ast_\sigma true$ is a needed narrowing
derivation, then $\sigma$ is a solution for $e$.
\item (Completeness) For each constructor substitution $\sigma$ that
is a solution of $e$, there exists a needed narrowing derivation
$e \leadsto^\ast_{\sigma'} true$ with $\sigma' \leq \sigma ~[\Var(e)]$.
\item (Minimality) If $e \leadsto^\ast_\sigma true$ and
$e \leadsto^\ast_{\sigma'} true$
are two distinct needed narrowing derivations,
then $\sigma$ and $\sigma'$ are independent on $\Var(e)$.
\end{enumerate}
\end{theorem}
An important advantage of functional logic languages
in comparison to pure logic languages
is their improved operational behavior by avoiding
non-deterministic computation steps.
One reason for that is a demand-driven
computation strategy which can avoid the evaluation
of potential non-deterministic expressions.
For instance, consider the rules in
Examples~\ref{example-leq} and~\ref{example-leqadd}
and the term \pr{0 \sleq X+X}.
Needed narrowing evaluates this term by one deterministic
step to \pr{true}. In an equivalent logic program,
this nested term must be flattened into a conjunction of two
predicate calls, like \pr{+(X,X,Z) \land \mathord{\sleq}(0,Z)},
which causes a non-deterministic computation due to the
predicate call \pr{+(X,X,Z)}.\footnote{Such non-deterministic computations
could be avoided using Prolog systems with coroutining
which allow the suspension of some non-deterministic computations,
but then we are faced with the problem of floundering
and incompleteness.}
Another reason for the improved operational behavior
of functional logic languages is the ability of
particular evaluation strategies (like needed narrowing
or parallel narrowing \cite{AEH97}) to evaluate ground terms
in a completely deterministic way, which is important to ensure
an efficient implementation of purely functional evaluations.
This property, which is obvious by the definition of needed narrowing,
is formally stated in the following proposition. For this purpose,
we call a term $t$ \emph{deterministically evaluable}
(w.r.t.\ needed narrowing) if each step in a narrowing
derivation issuing from $t$ is deterministic.
A term $t$ \emph{deterministically normalizes} to a constructor
term  $c$
(w.r.t.\ needed narrowing) if $t$ is deterministically evaluable
and there is a needed narrowing derivation $t \leadsto_{id}^\ast c$
(i.e., $c$ is the normal form of $t$).

\begin{proposition}\label{prop-nn-normalizing}
Let $\Rc$ be an inductively sequential program and $t$ be a term.
\vspace{-.2cm}
\begin{enumerate}
\item If $t \leadsto^\ast_{id} c$ is a needed narrowing
derivation, then $t$ deterministically normalizes to $c$.
\item If $t$ is ground, then $t$ is deterministically evaluable.
\end{enumerate}
\end{proposition}

\section{Lazy Narrowing and Uniform Programs}\label{sec-ln}

One of the main objectives of this work is to clarify the
relation between the definition of a PE scheme based on needed narrowing 
and a previous PE method based on lazy narrowing
\cite{AFJV97}.
In order to show the improvements obtained by using needed narrowing 
to perform partial computations, we first provide a brief review of the 
lazy narrowing strategy in this section.

Lazy narrowing reduces expressions at outermost narrowable
positions. Narrowing at inner positions is performed only if it is
demanded (by the pattern in the lhs of some rule). In the
following, we specify a lazy narrowing strategy which is similar to
\cite{MR92}.

The following definitions are necessary for our formalization of lazy
narrowing. A \emph{linear unification problem} is a pair of terms:
$\delta= \l f(\ol{d_n}),f(\ol{t_n})\r$,
where $f(\ol{d_n})$ and  $f(\ol{t_n})$ do not share variables,
and $f(\ol{d_n})$ is a linear pattern.
Linear unification {\sf LU}($\delta$) can either succeed, fail or suspend,
delivering $({\tt Succ}, \sigma)$,
$({\tt Fail}, \emptyset)$ or
$({\tt Demand}, P)$,
respectively, where $P$ is the set of \emph{demanded positions}
which require further evaluation; details can be found in \cite{AFJV97}.

We define the lazy narrowing strategy in the following definition.
Roughly speaking, the set-valued function
$\lns(t)$ returns the set of
triples $ (p, R, \sigma)$ such that $p$ is a
demanded position of $t$ which can be narrowed by the rule $R$ with
substitution $\sigma$
(where $\sigma$ is a most general unifier of $t|_p$ and the left-hand
side of $R$).
We assume the rules of $\Rc$
to be numbered with $R_1, \ldots,R_m$.

\begin{definition}[lazy narrowing strategy]\label{LNS} \mbox{}\\[-5ex]
\[ \hspace{-5ex}
\begin{array}{lll}
\lns(t) & = &
 \bigcup_{k=1}^{m} \lambda\_(t,\toppos,k) \\[0.15cm]
\lambda\_(t,p,k) & = & \mbox{if } root(l_k) = root(t|_p)
\mbox{ then} \\[0.1cm]
& & ~~~\mbox{case {\em {\sf LU}}}(\tuple{l_{k},t{}_{| p}})
                           \mbox{ of }
          \left\{ \begin{array}{ll}
           (\mbox{\tt Succ}, \sigma): & \{(p, R_{k}, \sigma)\} \\
           (\mbox{\tt Fail}, \emptyset): & \emptyset\\
           (\mbox{\tt Demand}, P): & \bigcup_{q \in P}
           \bigcup_{k=1}^{m} \lambda\_(t, p.q,k)
         \end{array} \right. \\
& &  \mbox{else } \emptyset
\end{array} \]
where $R_{k} = (l_k \rightarrow r_k) \mbox{ is a
(renamed apart) rule of } {\cal
R}$.
\end{definition}

\begin{example}\label{example-leq-lazy}
Consider the rules for ``$\sleq$'' and ``$+$''
in Examples~\ref{example-leq} and~\ref{example-leqadd}.
Then lazy narrowing evaluates the term $\tt X \sleq X+X$ by
applying a narrowing step at the top (with the first rule for ``$\sleq$'')
or by applying a narrowing step to the second argument $\tt X+X$
since this is demanded by the second and third rule for ``$\sleq$''.
Thus, there are three lazy narrowing steps:
\[
\begin{array}{lll}
\tt X \sleq X+X & \tt\leadsto_{\{X \mapsto 0\}}    & \tt true \\[1ex]
\tt X \sleq X+X & \tt\leadsto_{\{X \mapsto 0\}}    & \tt 0 \sleq 0 \\[1ex]
\tt X \sleq X+X & \tt\leadsto_{\{X \mapsto s(M)\}} & \tt s(M) \sleq s(M+s(M))
\end{array}
\]
Note that the second lazy narrowing step is in some sense superfluous
since it also yields the final value \pr{true} with the same
binding as the first step. The avoidance of such superfluous steps
by using needed narrowing
will have a positive impact on the PE process,
as we will see later.
\end{example}
In orthogonal programs, lazy narrowing is complete
w.r.t.\ strict equality and constructor substitutions:

\begin{proposition}[\citeNP{MR92}] \label{lazy-comple}
Let $\Rc$ be an orthogonal program, $e$ an equation, and $\sigma$ a
constructor substitution that is a solution for $e$.
Then there is a lazy narrowing derivation $e \leadsto^\ast_{\sigma'} true$
such that $\sigma' \leq \sigma ~[\Var(e)]$.
\end{proposition}
Thus, lazy narrowing is complete for a larger class of programs
than needed narrowing (since inductively sequential programs are always
orthogonal), but it may have a worse behavior than needed
narrowing (see Example~\ref{example-leq-lazy}).
Nevertheless, the idea of needed narrowing can also be extended
to almost orthogonal programs \cite{AEH97},
but then the optimality properties are lost.
There exists a class of programs where the superfluous steps
of lazy narrowing are avoided, since lazy narrowing and needed
narrowing coincide on this class. These are the \emph{uniform}
programs \cite{Zar97} which are inductively sequential programs
where at most one constructor occurs in the left-hand side
of each rule. A program is \emph{uniform} 
if each function $f$ is
defined by one rule $f(\ol{x_n}) \rightarrow r$ or the left-hand side
of every rule $R_i$ defining $f$ is left-linear and has
the form $f(\ol{x_k},c_i(\ol{y_{n_i}}),\ol{z_m})$,
where the constructors $c_i$ are distinct in different rules.
Note that uniform programs are orthogonal.
In the latter case, an evaluation of a call to $f$ demands
its $(k+1)$-th argument. A different definition of uniform
programs can be found in \cite{KLMR90b}.

There is a simple mapping $\cU$ from inductively sequential
into uniform programs which is based on flattening nested patterns,
see \cite{Zar97}.
For instance, if $\cR$ is the program in Example~\ref{example-leq},
then $\cU(\cR)$ consists of the rules
\[
\begin{array}{r@{~~\rightarrow~~}l}
\tt 0    \sleq N & \tt true \\
\tt s(M) \sleq N & \tt M \sleq' N
\end{array}
\qquad\qquad\qquad
\begin{array}{l@{~~\rightarrow~~}l}
\tt M \sleq' 0     & \tt false \\
\tt M \sleq' s(N1) & \tt M \sleq  N1
\end{array}
\]
where $\sleq'$ is a new function symbol.

The following theorem 
states a correspondence between needed
narrowing derivations using the original program and lazy
narrowing derivations in the transformed uniform program. For a
more detailed comparison between needed narrowing and lazy
narrowing, we refer to \cite{AFJV02}.

\begin{theorem}[\citeNP{Zar97}] \label{theo-zar97}
Let $\cR$ be an inductively sequential program, $\cU(\cR)$
the transformed uniform program, and $t$ an operation-rooted term.
Then there exists a needed narrowing derivation $t \leadsto^\ast_\sigma s$
w.r.t.\ $\Rc$ to a constructor root-stable form $s$ iff there exists
a lazy narrowing derivation $t \leadsto^\ast_\sigma s$ w.r.t.\ $\cU(\cR)$.
\end{theorem}

\section{Partial Evaluation with Needed Narrowing}\label{sec-nnpe}

In this section, we introduce the basic notions of PE in (lazy) 
functional logic programming. Then, we analyze the fundamental 
properties of PE based on needed narrowing and establish 
the relation with PE based on lazy narrowing.

Partial evaluation is a semantics-based program optimization
technique which has been investigated within different programming
paradigms and applied to a wide variety of languages. The first
PE framework for functional logic programs has been defined by 
\cite{AFV98}. In this framework, narrowing (the
standard operational semantics of integrated languages) is used to
drive the PE process; similarly to partial deduction,
specialized program rules are constructed from
narrowing derivations using the notion of \emph{resultant}. 
In the following, $s \leadsto^{+}_\sigma t$ denotes a narrowing
derivation with at least one narrowing step.

\begin{definition}[resultant]
Let $\cR$ be a TRS and $s$ be a term.
Given a narrowing  derivation
$s \leadsto^{+}_{\sigma} t$, its associated resultant
is the rewrite rule $\sigma(s) \rightarrow t$.
\end{definition}
Note that, whenever the specialized call $s$ is not a linear pattern,
the left-hand sides of resultants may not be linear patterns
either and hence resultants may not be program rules:

\begin{example}
Consider the following inductively sequential program:
\[
\begin{array}{rcl}
\tt double(X) & \to & \tt X+X \\
\tt 0+N & \to & \tt N \\
\tt s(M)+N & \to & \tt s(M+N) \\
\end{array}
\]
Given the term $\tt double(W)+W$ and the following needed narrowing
derivation (the selected redex is underlined at each
narrowing step):
\[
\tt \underline{double(W)}+W ~\leadsto_{id}~ (\underline{W+W})+W
~\leadsto_{\{W \mapsto s(M)\}}~
 s(M+s(M))+s(M)
\]
we compute the associated resultant:
\[
\tt double(s(M))+s(M) ~\to~ s(M+s(M))+s(M)
\]
This resultant is not a legal program rule since its left-hand side
contains nested defined function symbols (``\pr{+}'' and
``\pr{double}'') as well as multiple occurrences of the same
variable.
\end{example}
In order to produce legal program rules, we introduce a post-processing
of renaming which not only eliminates redundant
structures but also obtains \emph{independent} specializations in
the sense of \cite{LS91}. Furthermore, it is also 
necessary for the correctness of
the PE transformation. Roughly speaking, independence ensures that
the different specializations for the same function definition are
correctly distinguished, which is crucial for polyvariant
specialization.

The \emph{(pre--)partial evaluation} of a term $s$ is obtained by
constructing a (possibly incomplete) narrowing tree for $s$
and then extracting the
specialized definitions (the resultants) from the
non--failing, root--to--leaf paths of the tree.

\begin{definition}[pre--partial evaluation]
Let $\cR$ be a TRS and $s$ a term. Let ${\sf T}$ be a finite (possibly
incomplete) narrowing tree for $s$ in $\cR$ such that
no constructor root-stable term in the tree has been narrowed.
Let $\ol{t_{n}}$ be the terms in the non-failing leaves of ${\sf T}$. Then,
the set of resultants $\{ \sigma_{i}(s) \to t_{i} \mid i=1,\ldots,n\}$
for the narrowing sequences $\{ s
\leadsto^{+}_{\sigma_{i}} t_{i} \mid i = 1,\ldots,n \}$ is called
a pre--partial evaluation of $s$ in $\cR$.

The pre--partial evaluation of a set of terms $S$ in $\cR$ is defined as
the union of the pre--partial evaluations for the terms of $S$  in $\cR$.
\end{definition}

\begin{example} \label{app}
Consider the following function {\tt append} to concatenate two
lists (here we use ``{\tt nil}'' and ``{\tt :}'' as constructors of 
lists):
\[
\begin{array}{l}
\tt append(nil,Y_s)  \rightarrow  \tt Y_s \\
\tt append(X:X_s,Y_s)  \rightarrow  \tt X:append(X_s,Y_s)\\
\end{array}
\]
together with the set of calls
$S = \{ {\tt append(append(X_s,Y_s),Z_s), ~append(X_s,Y_s)} \}$.
Given the needed narrowing trees of Figure~\ref{nn-trees},
\begin{figure}[t]
{\tt    \setlength{\unitlength}{0.87pt}
\begin{center}
$\fbox{
\begin{picture}(350,186)(10,10)
\thinlines
              \put(107,145){{\scriptsize  $\{ {\rm X_s \mapsto nil} \}$}}
              \put(246,145){{\scriptsize  $\{ {\rm X_s \mapsto X':X'_s } \}$}}
              \put(274,113){\line(0,-1){14}}
              \put(155,171){\line(3,-1){120}}
              \put(154,171){\line(-3,-1){120}}
              \put(230,92){{${\tt X':append(\underline{append(X'_s,Y_s)},Z_s)}$}}
              \put(230,120){{${\tt append(X':append(X'_s,Y_s),Z_s)}$}}
              \put(10,120){{${\tt \underline{append(Y_s,Z_s)}}$}}
              \put(85,182){{${\tt append(\underline{append(X_s,Y_s)},Z_s)}$}}

              \put(107,45){{\scriptsize  $\{ {\rm Y_s \mapsto nil} \}$}}
              \put(246,45){{\scriptsize  $\{ {\rm Y_s \mapsto Y':Y'_s} \}$}}
              \put(155,71){\line(3,-1){120}}
              \put(154,71){\line(-3,-1){120}}
              \put(244,20){{${\tt Y':\underline{append(Y'_s,Z_s)}}$}}
              \put(30,20){{${\tt Z_s}$}}
              \put(120,82){{${\tt \underline{append(Y_s,Z_s)}}$}}
\end{picture}
}$
\end{center}}
\caption{Needed Narrowing trees for $\tt append(append(X_s,Y_s),Z_s)$ and
$\tt append(X_s,Y_s)$.} \label{nn-trees}
\end{figure}
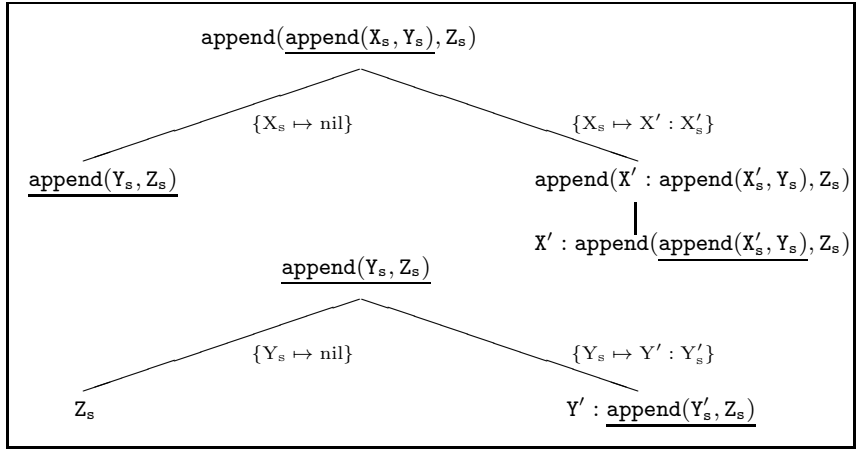
the associated pre--partial evaluation of $S$
in $\cR$ is as follows:
\[
\begin{array}{l}
\tt append(append(nil,Y_s),Z_s) \rightarrow append(Y_s,Z_s) \\
\tt append(append(X:X_s,Y_s),Z_s) \rightarrow X:append(append(X_s,Y_s),Z_s) \\
\tt append(nil,Z_s) \rightarrow Z_s \\
\tt append(Y:Y_s,Z_s) \rightarrow Y:append(Y_s,Z_s)
\end{array}
\]
\end{example}
The following example illustrates that the restriction not to evaluate 
terms in constructor root-stable form cannot be dropped.

\begin{example} \label{hnf-unsafe}
Consider the following program $\cR$:
\[
\begin{array}{rcl}
\tt f(0) & \rightarrow & \tt 0 \\
\tt g(X) & \rightarrow & \tt s(f(X)) \\
\tt h(s(X)) & \rightarrow & \tt s(0) \\
\end{array}
\]
together with the set of calls $S = \tt \{g(X),h(X)\}$. Given the 
needed narrowing derivations:
\[
\begin{array}{l}
\tt \underline{g(X)} ~\leadsto_{id}~ s(\underline{f(X)}) 
~\leadsto_{\{X\mapsto 0\}}~ s(0) \\
\tt \underline{h(X)} ~\leadsto_{\{X\mapsto s(Y)\}}~ s(0)
\end{array}
\]
a pre--partial evaluation of $S$ in $\cR$ is the following 
program $\cR'$:
\[
\begin{array}{rcl}
\tt g(0) & \rightarrow & \tt s(0) \\
\tt h(s(X)) & \rightarrow & \tt s(0) \\
\end{array}
\]
Now, the equation $\tt h(g(s(0))) \approx X$ has the following successful
needed narrowing derivation in $\cR$:
\[
\tt h(\underline{g(s(0))}) \approx X ~\leadsto_{id}~ \underline{h(s(f(s(0))))}
   \approx X ~\leadsto_{id}~ s(0) \approx X ~\leadsto^{\ast}_{\{X\mapsto s(0)\}}
   true
\]
whereas it fails in the specialized program $\cR'$.
\end{example}
The problem shown in the above example is due to the \emph{backpropagation}
of bindings to the left-hand sides of resultants: within a lazy context, the
instantiation of the left-hand sides of resultants with bindings which
come from the evaluation of terms in constructor root-stable form
may incorrectly restrict the domain of functions (e.g., 
function ``\pr{g}'' above).

A recursive \emph{closedness} condition, which
guarantees that each call which
might occur during the execution
of the resulting program is covered by some program rule, is
formalized by inductively checking that the different calls
in the rules are sufficiently covered
by the specialized functions. For instance, a function call like $\tt
s(X)+Y$ cannot be considered \emph{closed} w.r.t.\ the set of calls
$\tt \{0+Y,s(0)+Y\}$.

Informally, a term $t$ rooted by a defined function symbol
is closed w.r.t.\ a set of calls $S$, if it is an
instance of a term of $S$ and the terms in the matching
substitution are recursively closed by $S$.

\begin{definition}[closedness] \label{closedness}
Let $S$ be a finite set of terms. We say that a term
$t$ is $S$-closed if $closed(S,t)$ holds, where the predicate $closed$ is
defined inductively as follows:
\[ \hspace{-6ex}
\begin{array}{@{}l}
closed(S,t) \: \Leftrightarrow
\left\{
\begin{array}{ll}
true & \mbox{if } t  \in \cX \\
closed(S,t_1)  \wedge  \ldots  \wedge  closed(S,t_n) & \mbox{if } t =
c(\ol{t_n}), 
~c \in \cC^\ast, ~ n\geq 0\\
\bigwedge_{x\mapsto t' \in
\theta}
closed(S,t') & \mbox{if } \exists s \in S 
\mbox{ such that } \theta(s) = t \\
& \mbox{ for some substitution $\theta$}
\end{array} \right.
\end{array}
\]
where $\cC^\ast = (\cC \cup\{\equ,\wedge\})$.

We say that a set of terms $T$ is $S$-closed, written $closed(S,T)$, if
$closed(S,t)$ holds for all $t \in T$, and we say that a TRS $\cR$ is
$S$-closed if $closed(S,\cR_{calls})$ holds.
Here we denote by $\cR_{calls}$ the set of the right-hand sides of the rules
in $\cR$.
\end{definition}
For instance, the pre--partial evaluation of Example~\ref{app} is closed
w.r.t.\ the set of partially evaluated calls
$\tt \{append(append(X_s,Y_s),Z_s), ~append(X_s,Y_s)\}$.

According to the (non-deterministic)
definition above, an expression rooted by a ``primitive''
function symbol, such as a conjunction $t_{1} \wedge t_{2}$
or an equation  $t_{1} \equ t_{2}$,
can be proved closed w.r.t.\ $S$
either by checking
that $t_{1}$ and $t_{2}$ are $S$-closed or by testing whether
the conjunction (equation)
 is an instance of a call in $S$
(followed by an inductive test of the subterms).
This is useful when we
are not interested in specializing
complex expressions (like conjunctions or equations)
but we still want to run them after specialization.
Note that this is safe since we consider that the  rules which define the
primitive functions ``$\equ$'' and ``$\wedge$'' 
are automatically added to each program by existing programming
environments, hence
calls to these  symbols are steadily covered in the  specialized program.
A general technique for dealing with primitive symbols
which deterministically splits
terms before testing them for closedness 
can be found in \cite{AAFJV98}.

In general, given a call $s$ and a program ${\cal R}$, there exists an infinite
number of different pre--partial evaluations of $s$ in ${\cal R}$. A fixed rule
for generating resultants called an \emph{unfolding rule} is assumed,
which determines the expressions to be narrowed
(by using a fixed narrowing strategy)
and which decides how to stop the construction of  narrowing trees;
see \cite{AAFJV98,AFV98,AHV02} for the definition of concrete unfolding
rules.

In the following, we denote by pre--NN--PE and pre--LN--PE
the sets of resultants computed for $S$ in $\cR$
by considering an unfolding rule
which constructs  finite needed and lazy narrowing trees, respectively.
We will use the acronyms NN--PE and LN--PE
for the renamed  rules which will result from the
corresponding post-processing of \emph{renaming}.
The idea behind this
transformation is that, for any call (which is closed w.r.t.\ the 
considered set of calls),
the answers computed for this call in the original program
and the answers computed for the renamed call  in the
specialized, renamed program do coincide.
In particular, in order to define a partial evaluator
based on needed narrowing
and to ensure that the resulting program is inductively
sequential whenever the source program is, we have to
make sure that the set of specialized terms (after renaming)
contains only linear
patterns with distinct root symbols.
This can be ensured by introducing a new function symbol
for each specialized term and then replacing
each call in the specialized program by a call to the corresponding
renamed function. In particular,
the left-hand sides of the specialized
program (which are constructor instances of the specialized terms)
are replaced by instances of the corresponding new linear patterns
through renaming.

\begin{definition}[independent renaming] \label{ren} An
independent renaming $\rho$ for a set of terms $S$ is a
mapping  from terms to terms defined as
follows: for $s\in S$,
$\rho(s)={f_{s}(\ol{x_{n}})}$,
where $\ol{x_{n}}$ are the distinct variables in $s$ in the
left-to-right ordering and $f_{s}$ is a new function symbol,
which does not occur in $\Rc$ or $S$ and
is different from the root symbol of any other $\rho(s')$,
with $s'\in S$ and $~s'\neq s$. We also denote by $\rho(S)$ 
the set $S'=\{\rho(s)\mid s\in S\}$.
\end{definition}

\begin{example} \label{iren_app}
Consider the set
$S = \{ {\tt append(append(X_s,Y_s),Z_s), ~append(X_s,Y_s)} \}$. The
following mapping:
\[ \hspace{-5ex}
\rho = \{ \tt append(X_s,Y_s) \mapsto app(X_s,Y_s), ~
\tt append(append(X_s,Y_s),Z_{s}) \mapsto dapp(X_s,Y_s,Z_s) ~ \}
\]
is an independent renaming for $S$.
\end{example}
While independent renamings suffice to rename the left-hand sides
of resultants (since they are constructor instances of the specialized
calls), the right-hand sides are renamed by means of the auxiliary function
$ren_{\rho}$, which \emph{recursively} replaces
each call in the given expression
by a call to the corresponding renamed function (according to $\rho$).

\begin{definition}[renaming function] \label{ren2}
Let $S$ be a finite set of terms and $\rho$ an independent
renaming of $S$. Given a term $t$, the non-deterministic function
$ren_{\rho}$ is defined as follows:
\[
ren_{\rho}(t) = \left\{ \begin{array}{ll}
t  &  \mbox{if } t \in \cX \\
c(\ol{ren_{\rho}(t_{n})}) & \mbox{if } t = c(\ol{t_n}),~
                           c \in  \cC^\ast, \mbox{ and }~ n \geq 0 \\
\theta'(\rho(s))  & \mbox{if } \exists \theta, \exists s \in S \mbox{ such that }
t = \theta(s) \mbox{ and } \\
& ~~~~ \theta' = \{ x \mapsto ren_{\rho}(\theta(x))
\mid x \in \sdom(\theta) \} \\
t & \mbox{otherwise}
\end{array} \right.
\]
where $\cC^\ast = ({\cal C} \cup \{\equ,\land\})$.
\end{definition}
Similarly to the test for closedness, an
equation $s \equ t$ can be (non-deterministically) renamed either by
independently
renaming $s$ and $t$ or by replacing the considered equation
by a call to the corresponding new, renamed
function (when the equation is an instance of some specialized
call in $S$). Note also that the renaming function is a \emph{total}
function: if an operation-rooted term $t$ is
not an instance of any term in $S$ (which can occur if $t$ is not
$S$-closed), the function $ren_{\rho}(t)$ returns $t$ itself
(i.e., term $t$ is not renamed).

The notion of partial evaluation can be formally defined  as follows.

\begin{definition}[partial evaluation] \label{p-p-r}
Let $\Rc$ be a TRS, $S$ a finite set of terms and
 $\Rc'$ a pre--partial
evaluation of $\cal R$ w.r.t.\ $S$.
Let $\rho$ be  an independent renaming
of $S$. We define the partial evaluation
$\Rc''$ of $\cR$ w.r.t.\ $S$ (under $\rho$)
as follows:
\[ \Rc''=
{\bigcup_{s \in S}} \{ \theta(\rho(s)) \rightarrow ren_{\rho}(r)
\mid \theta(s) \rightarrow r \in \Rc' \mbox{ is a resultant for $s$ in $\Rc$} \}
\]
\end{definition}
We now illustrate these definitions with an example.

\begin{example} \label{ppren_app}
Let us consider the
program {\tt append} and the set of terms $S$ of Example~\ref{app}, together
with the independent renaming $\rho$ of Example~\ref{iren_app}.
A partial evaluation $\Rc'$ of $\Rc$ w.r.t.\ $S$ (under $\rho$)
is:
\[
\begin{array}{rcl}
\tt dapp(nil,Y_s,Z_s) & \rightarrow & \tt  app(Y_s,Z_s) \\
\tt dapp(X:X_s,Y_s,Z_s) & \rightarrow & \tt X:dapp(X_s,Y_s,Z_s) \\
\tt app(nil,Y_s) & \rightarrow & \tt  Y_s \\
\tt app(X:X_s,Y_s) & \rightarrow & \tt  X:app(X_s,Y_s)
\end{array}
\]
\end{example}
Note that, for a given renaming $\rho$, the renamed form of
a program $\Rc$ may depend on the strategy which selects the term from
$\rho(S)$ which
is used to rename a given call $t$ in $\Rc$
(e.g., $\tt append(append(X_s,Y_s),Z_{s})$), since there may exist, in general,
more than one term in $S$ that covers the call $t$.
Some potential specialization might be lost due to an inconvenient choice.
Appropriate heuristics which are able to produce the best potential 
specialization have been introduced in the implementation of the partial 
evaluator described in \cite{AHV02}.

The correctness of LN-PE is stated in \cite{AAFJV98,AFJV97}. 
It is important to clarify that, even if the methodology for
narrowing-driven PE in \cite{AFV98} is
parametric w.r.t.\ the narrowing strategy, this framework
only ensures that: 
\begin{itemize}
\item partially evaluated programs are \emph{closed} w.r.t.\ the
set of partially evaluated calls---which is necessary, although does 
not suffice, to guarantee the completeness of the transformation---, 
and
\item the PE process always terminates.
\end{itemize}
In particular, the
correctness of the PE transformation cannot be proved in a way
independent of the narrowing strategy. These results are by
their nature highly dependent on the concrete strategy which is
considered, as it is known that different narrowing strategies
have quite different semantic properties. In fact, the use of a
lazy evaluation strategy imposes some additional restrictions on
PE, such as the use of ``strict equality'', the requirement not
to evaluate terms in constructor root-stable form during PE, or the need for
an additional post-processing of renaming. All these additional
requirements are essential to ensure the correctness of the
transformation and were not present in the original framework of
\cite{AFV98,AFV98b}, where 
correctness is only proved for an
eager narrowing strategy. Therefore, it was necessary to
develop a new theory for PE based on lazy narrowing
as a separate work \cite{AFJV97}, which is now overcome by
the needed narrowing methodology formalized in this article.

The following lemma shows that any PE based on needed
narrowing can also be obtained (but possibly with more steps)
by PE of the transformed uniform program
based on lazy narrowing. This means that, in some sense,
the specializations computed by a partial evaluator based on
needed narrowing cannot be worse than the specializations
computed by a partial evaluator based on lazy narrowing.
On the other hand, we will also show later
that there are cases where a LN-PE is worse than a NN-PE
for the same original program.

\begin{lemma}\label{LN-NN}
Let $\Rc$ be an inductively sequential program, $\Rc_u = \cU(\Rc)$ the
corresponding uniform program, and $S$ a finite set of operation-rooted
terms.
If $\Rc'$ is an NN-PE of $S$ in $\Rc$, then $\Rc'$ is also an LN-PE
of $S$ in $\Rc_u$.
\end{lemma}
\begin{proof}
Since the final renaming applied in the partial evaluation
of a program does not depend on the narrowing strategy
used during the pre-partial evaluation,
it suffices to show that each resultant w.r.t.\ needed narrowing in $\Rc$
corresponds to a resultant w.r.t.\ lazy narrowing in $\Rc_u$.
Due to the definition of a resultant, each rule in the pre-partial
evaluation w.r.t.\ needed narrowing in $\Rc$ has the form
\[
\sigma(t) \to s
\]
where $t \in S$ and $t \leadsto_\sigma^{+} s $ is a needed narrowing
derivation w.r.t.\ $\Rc$. By Theorem~\ref{theo-zar97},
there exists a lazy narrowing derivation
$t \leadsto_\sigma^{+} s$ w.r.t.\ $\Rc_u$
which has the same answer and result
(note that Theorem~\ref{theo-zar97} states this property only
for derivations into constructor-rooted terms, but it also holds
in the direction used here for arbitrary needed narrowing derivations
since each needed narrowing step corresponds to a sequence of
lazy narrowing steps w.r.t.\ the transformed uniform programs,
which can be seen by the proof of this theorem).
Thus, $\sigma(t) \to s$ is a resultant of this lazy narrowing
derivation w.r.t.\ $\Rc_u$.
\end{proof}
The following theorem states an important property of PE based on
needed narrowing: if the input program is inductively
sequential, then the partially evaluated program is also inductively
sequential and, thus, we can also apply the needed
narrowing strategy to evaluate calls in the specialized program.
The proof of this theorem can be found in
\ref{app-nn-pe-is}. An extension of this theorem---although it relies
on the result below regarding the unfolding transformation---in the 
context of a more general fold/unfold framework can be found in \cite{AFMV04}.
\begin{theorem}\label{theo-nn-pe-is-terms}
Let $\Rc$ be an inductively sequential program and $S$ a finite
set of operation-rooted terms. Then each NN-PE of $\Rc$ w.r.t.\ $S$
is inductively sequential.
\end{theorem}
The following example reveals that,
when we consider lazy narrowing, the LN-PE of a uniform program
w.r.t.\ a linear pattern may not be uniform.

\begin{example}
Let $\cR$ be the following uniform program:
\[ \begin{array}{rcl}
\tt f(X,b) & \rightarrow & \tt g(X) \\
\tt g(a) & \rightarrow & \tt a
\end{array} \]
and $t = {\tt f(X,Y)}$ and $\rho(t) = {\tt f2(X,Y)}$.
Then a LN-PE $\cR'$ of $t$ in $\cR$ (under $\rho$) is
\[ 
\begin{array}{rcl}
{\tt f2(a,b)} & \rightarrow & {\tt a}
\end{array} 
\]
which is not uniform.
\end{example}
The residual program $\cR'$ in the example
above is inductively sequential. This raises the question whether
the LN-PE of a uniform program is always inductively sequential.
Corollary \ref{preser-is} will positively answer this question.

\begin{corollary}\label{preser-is}
Let $\Rc$ be a uniform program and $S$ a finite set of operation-rooted
terms. If $\Rc'$ is a LN-PE of $S$ in $\Rc$, then $\Rc'$
is inductively sequential.
\end{corollary}
\begin{proof}
Since a uniform program is inductively sequential and lazy narrowing steps
w.r.t.\ uniform programs are also needed narrowing steps
(cf.\ proof of Theorem~\ref{theo-zar97}),
the proposition is a direct consequence of
Theorem~\ref{theo-nn-pe-is-terms}.
\end{proof}
The uniformity condition in Corollary \ref{preser-is} cannot be
weakened to inductive sequentiality when LN-PEs are
considered, as demonstrated by the
following counterexample.

\begin{example}
Let $\cR$ be the following inductively sequential program:
\[ \begin{array}{rclcrcl}
\tt f(a,a,a) & \rightarrow & \tt b &~~~~& \tt h(a,b,X) & \rightarrow & \tt b \\
\tt f(b,b,X) & \rightarrow & \tt b && \tt h(e,X,k) & \rightarrow & \tt b \\
\tt g(a,b,X) & \rightarrow & \tt b && \tt i(X,c,d) & \rightarrow & \tt b\\
\tt g(X,c,d) & \rightarrow & \tt b && \tt i(e,X,k) & \rightarrow & \tt b
\end{array} \]
Let $t= {\tt f(g(X,Y,Z),h(X,Y,Z),i(X,Y,Z))} \in S$ and $\rho$ be a renaming
such that $\rho(t) = \tt f3(X,Y,Z)$.
Then, every LN-PE $\:\cR'$ of $S$ in $\cR$ (considering depth-2 lazy narrowing trees
to construct the resultants)  contains the rules:
\[ \begin{array}{rcl}
\tt f3(a,b,X) & \rightarrow & \cdots \\
\tt f3(e,X,k) & \rightarrow & \cdots \\
\tt f3(X,c,d) & \rightarrow & \cdots \\
\end{array} \]
and thus $\cR'$ is not inductively sequential.
\end{example}
One of the main factors affecting the quality of a PE is the treatment
of choice points \cite{LB02,Gal93}. The following examples illustrate
the different way in which NN-PE and LN-PE
``compile-in''  choice points during unfolding, which
is crucial to performance
since a poor control choice during
the construction of the computation trees  can
inadvertently introduce extra computation into a program.

\begin{example}
Consider again  the rules of Example~\ref{example-leqadd} and
the input term $\tt X \sleq X+Y$.
The computed LN-PE is as follows:
\[
\begin{array}{r@{~~\rightarrow~~}l}
\tt leq2(0,N) & \tt true \\
\tt leq2(0,N') & \tt true \\
\tt leq2(s(M),N) & \tt leq2(M,N)
\end{array}
\]
where the renamed initial term is \pr{leq2(X,Y)}.
The redundancy of lazy narrowing has the effect that the first
two rules of the specialized program are identical (up to renaming).
In contrast, a better specialization---without generating redundant 
rules---is obtained by PE based on needed narrowing,
since the NN-PE consists of the following rules:
\[
\begin{array}{r@{~~\rightarrow~~}l}
\tt leq2(0,N) & \tt true \\
\tt leq2(s(M),N) & \tt leq2(M,N)
\end{array}
\]
Note that a call-by-value partial evaluator based on innermost
narrowing \cite{AFV98} has an even worse behavior in this example
since it does not specialize the program at all.
\end{example}
In the example above, the superfluous rule in the LN-PE can be avoided by
removing duplicates in a post-processing step.
The next example shows that this is not always possible.

\begin{example} \label{ex-michael}
Lazy evaluation strategies are necessary if one wants to deal
with infinite data structures and possibly non-terminating
function calls. The following orthogonal program makes 
use of these features:
\[
\begin{array} {rclcrcl}
\tt f(0,0)  & \rightarrow  & \tt s(f(0,0)) &~~~~~~~~~&\tt g(0)   &\rightarrow   & \tt g(0) \\
\tt f(s(N),X) & \rightarrow & \tt s(f(N,X)) &~~~~~~~~~&
\tt h(s(X))  & \rightarrow   & \tt 0
\end{array}
\]
The specialization is initiated with the term \pr{h(f(X,g(Y)))}.
Note that this term reduces to \pr{0} if \pr{X} is bound to
\pr{s(\cdots)}, and it does not terminate if \pr{X} is bound to
\pr{0} due to the nonterminating evaluation of the second argument.
The NN-PE of this program perfectly reflects this behavior
(the renamed initial term is \pr{h2(X,Y)}):
\[
\begin{array}{rclcrcl}
\tt h0      &\rightarrow    & \tt h0 &~~~~~~~~~~~~~~~~~~&
\tt h2(0,0)    &\rightarrow    & \tt h0 \\
 ~~~~~~~~~~~~~       && &~~~~~~~~~~~~~~~~~~&\tt h2(s(X),Y)  &\rightarrow   & \tt 0
\end{array}
\]
On the other hand, the LN-PE of this program has a worse structure:
\[
\begin{array}{rclcrcl}
\tt h1(X)  &\rightarrow & \tt h1(X) &~~~~~~~~~~~~~&
\tt h2(X,0)   &\rightarrow   & \tt h1(X) \\
\tt h1(s(X))& \rightarrow & \tt 0 &~~~~~~~~~~~~~&\tt h2(s(X),Y)  &\rightarrow & \tt 0 \\
                && &~~~~~~~~~~~~~&\tt h2(s(X),0) &\rightarrow  & \tt 0
\end{array}
\]
\end{example}
The program specialized by LN-PE in the example above
is not inductively sequential (nor orthogonal),
in contrast to the original one. This does not only mean that
lazy and needed narrowing are not applicable to the specialized program but
also that the specialized program has a worse termination behavior than
the original one. For instance, consider the term \pr{h(f(s(0),g(0)))}.
The evaluation of this term has a finite derivation tree
w.r.t.\ lazy narrowing as well as needed narrowing in the original 
program. However, the renamed term \pr{h2(s(0),0)}
has a finite derivation tree w.r.t.\ the NN-PE but
an infinite derivation tree w.r.t.\ the LN-PE (using lazy narrowing);
the infinite branch is caused by the application of the rules
\pr{h2(X,0) \to h1(X)} and \pr{h1(X) \to h1(X)}.

This last example also shows that LN-PE can destroy the advantages 
of deterministic reduction of functional logic programs, which is
not possible using NN-PE. This is ensured by the
following theorem, which guarantees that a term which is
deterministically normalizable w.r.t.\ the original program
cannot cause a non-deterministic evaluation w.r.t.\ the specialized
program obtained by NN-PE.

\begin{theorem}
Let $\cR$ be an inductively sequential program,
$S$ a finite set of operation-rooted terms,
$\rho$ an
independent renaming of $S$, and $e$ an equation.
Let $\cR'$ be a NN-PE of $\cR$ w.r.t.\ $S$ (under $\rho$)
such that $\cR' \cup \{e'\}$ is $S'$-closed,
where
$e' = ren_{\rho}(e)$ and
$S' = \rho(S)$.
If $e$ deterministically normalizes to $true$ w.r.t.\ $\cR$,
then $e'$ deterministically normalizes to $true$ w.r.t.\ $\cR'$.
\end{theorem}
\begin{proof}
Since $e$ deterministically normalizes to $true$ w.r.t.\ $\cR$,
there is a needed narrowing derivation
$e \leadsto_{id}^\ast true$ in $\cR$.
By Theorem~\ref{theo-nnpe-strong-correct} (see below),
there is a needed narrowing derivation
$e' \leadsto_{\sigma}^\ast true$ in $\cR'$
with $\sigma = id ~[\Var(e)]$. This implies $\sigma = id$
by definition of needed narrowing.
Therefore, $e'$ deterministically normalizes to $true$ w.r.t.\ $\cR'$
by Proposition~\ref{prop-nn-normalizing}.
\end{proof}
This property of specialized programs is desirable and
important from an implementation point of view, since the
implementation of non-deterministic steps is an expensive
operation in logic-oriented languages. Moreover, additional
non-determinism in the specialized programs can result in
additional infinite derivations, as shown in Example
\ref{ex-michael}. This might have the effect that solutions are no
longer computable in a sequential implementation based on
backtracking. 
Essentially, deterministic computations are preserved thanks to the
use of needed narrowing over inductively sequential programs to
perform  partial computations. For instance, consider the function
``\pr{leq}'' of Example~\ref{example-leq}
together with the simple function ``\pr{foo}'':
\[
\begin{array}{r@{~~\rightarrow~~}l}
\tt foo(0) & \tt 0 \\
\end{array}
\]
Given a function call of the form $\tt X \sleq foo(Y)$, many narrowing
strategies (e.g., lazy narrowing) have two ways to
proceed: either by reducing the call to function ``\pr{\sleq}'' using
the first rule
\[
\tt X \sleq foo(Y) ~~\leadsto_{\{X \mapsto 0\}}~~true
\]
and by reducing the call to function ``\pr{foo}'' (which is demanded
by the second and third rules of ``\pr{\sleq}'')
\[
\tt X \sleq foo(Y) ~~\leadsto_{\{Y\mapsto 0\}}~~X \sleq 0\\
\]
Thus, their associated resultants are as follows:
\[
\begin{array}{r@{~~\rightarrow~~}l}
\tt 0 \sleq foo(Y) & \tt true\\
\tt X \sleq foo(0) & \tt X \sleq 0\\
\end{array}
\]
Now, given a call of the form $\tt 0 \sleq foo(Z)$, both resultants
are applicable but  the second one is clearly redundant.  Actually, the
second resultant is only meaningful to evaluate those calls whose
first argument is of the form $\tt s(\cdots)$, since only the second
and third rules of ``\pr{\sleq}'' demanded the evaluation of call {\tt foo(0)} 
that gave
rise to this resultant. The advantage of using needed narrowing is
that it applies some additional bindings so that this information is
made explicit in the computed resultants, e.g., the resultants
obtained by needed narrowing are
\[
\begin{array}{r@{~~\rightarrow~~}l}
\tt 0 \sleq foo(Y) & \tt true\\
\tt  \underline{s(Z)} \sleq foo(0) & \tt s(Z) \sleq 0\\
\end{array}
\]
thus avoiding the creation of additional non-determinism.  This
property is somehow related to the notion of \emph{perfect splits}
used in \cite{AG00,AG02,GK93} to guarantee that no computations are
neither lost nor added when constructing---by driving \cite{Tur86}, a
symbolic execution mechanism which shares many similarities with lazy
narrowing---the perfect process trees of (positive) supercompilation
\cite{SGJ93}.

Note that there is no counterpart of this property in the
partial deduction of logic programs, since the considered execution
mechanism (some variant of SLD-resolution) never demands---in a
\emph{don't-know} non-deterministic way---the evaluation of different
atoms of the same goal.

Finally, we state the strong correctness of NN-PE, which amounts to
the computational equivalence between the original and the
specialized programs (i.e., the fact that the two programs compute
exactly the same answers) for the considered goals.  The proof of this
theorem can be found in~\ref{sc-nn-pe}.

\begin{theorem}[strong correctness]\label{theo-nnpe-strong-correct}
Let $\cR$ be an inductively sequential program. Let $e$ be an equation,
$V \supseteq \var(e)$ a finite set of variables, $S$ a
finite set of operation-rooted terms, and $\rho$ an
independent renaming of $S$. Let $\cR'$ be a
NN-PE of $\cR$ w.r.t.\ $S$ (under $\rho$)
such that $\cR' \cup \{e'\}$ is $S'$-closed,
where $e' = ren_{\rho}(e)$ and $S' = \rho(S)$. Then, $e
\leadsto_{\sigma}^{\ast} true$ is a needed narrowing
derivation for $e$ in $\cR$ iff there exists a needed
narrowing derivation $e' \leadsto_{\sigma'}^{\ast} true$ in $\cR'$ such
that $(\sigma' = \sigma)~[V]$ (up to renaming).
\end{theorem}
It is worthwhile to note that the correctness of NN-PE cannot be 
derived from the correctness of LN-PE \cite{AFJV97}, since the 
preservation of inductive sequentiality (cf.\ 
Theorem~\ref{theo-nn-pe-is-terms}) is a crucial point in our 
proof scheme, and this property does not hold for LN-PE.

On the other hand, it is well-known that partial evaluation can be
defined within the fold/unfold framework \cite{PP96b} by using only
unfolding and a restricted form of folding. Hence the correctness of NN-PE 
could be derived from the correctness of a fold/unfold framework for
the transformation of functional logic programs based on needed
narrowing. However, the only framework of this kind in the literature
is \cite{AFMV99,AFMV04} and their proofs of correctness---regarding
the unfolding transformation---rely on the results  in this article. The
precise relation between partial evaluation and the fold/unfold
transformation---for lazy functional logic programs---can be found in
\cite{AFMV00}.

\section{Further Developments}\label{extensions}

In the previous sections, we introduced the theoretical basis for PE
in the context of lazy functional logic programming. 
Since the preliminary publication of these results, several extensions 
as well as concrete partial evaluators have been developed. 
In this section, we review some of these subsequent developments.

The computational model of modern declarative multi-paradigm 
languages, which integrate 
the most important features of functional, logic and 
concurrent programming, is based on a combination of two 
different operational principles: needed
narrowing and residuation \cite{Han97b}.
The \emph{residuation} principle is based on 
the idea of delaying
function calls until they are sufficiently instantiated 
for a deterministic evaluation by rewriting.
The particular mechanism (narrowing or residuation) is specified by 
\emph{evaluation annotations}:
deterministic functions are annotated as \emph{rigid} (which forces 
a delayed evaluation by rewriting), while non-deterministic functions 
are annotated as \emph{flexible} (which enables narrowing steps).

Although NN-PE is originally formulated for functional logic languages
based uniquely on needed narrowing, it is still possible to adapt it
to the use of distinct operational mechanisms. In fact, NN-PE has been
already adjusted to perform partial computations using the combined
operational semantics described above \cite{Alb01,AAHV99}.

On the other hand, NN-PE has also been
extended \cite{AHV02} in order 
to make it viable for defining partial evaluators for
\emph{practical} multi-paradigm functional logic 
languages like Curry \cite{CurryTR} or Toy \cite{LS99}.
When one considers a practical language, several 
extensions have to be considered, e.g., higher-order functions,
concurrent constraints, calls to external functions, etc.
In order to deal with these additional features, the underlying 
operational calculus becomes usually more complex.
As we mentioned earlier, an on-line partial evaluator normally includes an 
interpreter of the language \cite{CD93}. Then, as the 
operational semantics becomes more elaborated, the 
associated PE techniques become (more powerful but) also 
increasingly more complex. 
To avoid this problem, an approach successfully tested in other 
contexts \cite{Bon89,GK93,NPT96} is to consider the PE of
programs written in a maximally simplified programming language. 

\citeN{HP99} have introduced a \emph{flat} 
representation for functional logic programs in which
definitional trees are embedded 
in the rewrite rules by means of case expressions:

\begin{example}
Function ``$\sleq$'' of Example~\ref{example-leq}
can be written in the flat representation as follows:
\[ \hspace{-3ex}
\begin{array}{lllllllll}
\tt X \sleq Y = ~case\;\; X \;\; of & \tt \{ & \tt 0 & \tt \to & \tt true;\\
& \tt  & \tt s(X_1) & \tt \to&  \tt case\;\; Y \;\;of~ & \tt \{ & \tt 0 \to ~ false; \\
&&&&&&  \tt s(Y_1) \to ~X_1 \sleq Y_1 & \} & \}
\end{array}
\]
\end{example}
Two nice properties of the flat representation are that it provides 
more explicit control---hence the associated calculus is 
simpler than needed narrowing---and source programs can be 
automatically translated to the new representation. 
Moreover, it constitutes the basis of a 
recent proposal for an intermediate language, FlatCurry, 
used during the compilation of Curry programs \cite{AH00,AHMS01}.
A new PE scheme \cite{Alb01,AHV02} has been 
designed by considering such a \emph{flat} representation for 
functional logic programs.

However, the use of the standard semantics for flat programs---the LNT
calculus \cite{HP99}, which is equivalent to needed narrowing---at PE
time does not avoid the backpropagation of bindings when evaluating
terms in constructor root-stable form, which can be problematic within
a lazy context (see Example~\ref{hnf-unsafe}). In order to overcome
this problem, a \emph{residualizing} version of the standard semantics
is introduced: the RLNT calculus \cite{Alb01,AHV02b}.
Finally, since modern lazy functional logic languages can be automatically 
translated into this flat representation---which still contains all 
the necessary information about programs---the resulting technique is 
widely applicable. 

All these results laid the ground for the development of a partial
evaluation tool for Curry programs, which has been distributed with
the Portland Aachen Kiel Curry System \cite{PAKCS03} since April 2001.
Our partial evaluator constructs optimized, residual versions for
selected calls of the input program. These calls are annotated by means of
the function $\tt PEVAL$ which is equivalent to the identity function.
Let us show a typical session
with the partial evaluator.  Here we consider the optimization of a
program containing several calls to higher-order functions (since it
is common to use higher-order combinators such as \pr{map},
\pr{foldr}, etc.\ in Curry programs).  Although the use of such
functions makes programs concise, some overhead is introduced at run
time.  Hence, we apply our partial evaluator to optimize calls to
these functions.  As a concrete example, consider the following
(annotated) Curry program:\footnote{Here we follow the Curry syntax:
  both variables and functions (except for $\tt PEVAL$) start with
  lower case letters and function application is denoted by
  juxtaposition.}
\startprog
main xs ys = (PEVAL (map (iter (+1) 2) xs)) ++ ys\\[-2ex]
iter f n = if n==0 then f else iter (comp f f) (n-1)
comp f g x = f (g x)\\[-2ex]
bench = main [1..20000] []
\stopprog
stored in the file {\tt map\_iter.curry}.  Function \pr{comp} is a
higher-order function to compose two input functions, while \pr{iter}
composes a given function $2^{\tt n}$ times. Thus, given two input
lists, \pr{xs} and \pr{ys}, function \pr{main} adds $4$ to each
element of \pr{xs}---the annotated expression---and then concatenates
the result with the second list \pr{ys}. The built-in function
``\verb$++$'' denotes list concatenation in Curry (more details can be
found in \cite{CurryTR}). In order to measure the improvement achieved
by the process, we have also included the function \pr{bench} with a
simple call to function \pr{main}, where \verb$[1..20000]$ represents
a list from \verb$1$ to \verb$20000$.  First, we load the program into
PAKCS, turn on the time mode (to obtain the run time of computations),
and execute function \pr{bench}:
\startprog
prelude> :l map\_iter\\[-4ex]
...\\[-3ex]
{compiled /tmp/map\_iter.pl in module user, 620 msec 9888 bytes}
map\_iter> :set +time
map\_iter> bench
Runtime: 750 msec.
Result: [5,6,7,8,9,10,11,12,13,14,15,16,17,18,19,20,21,22,...]
\stopprog
Now, we run the partial evaluation tool and show the result of the
process:
\startprog
map\_iter> :peval\\[-4ex]
...\\[-3ex]
Writing specialized program into "map\_iter\_pe.flc"...
Loading partially evaluated program "map\_iter\_pe"...
map\_iter\_pe> :show
main xs ys = (map\_pe0 xs) ++ ys\\[-2ex]
iter f n = if n==0 then f else iter (comp f f) (n-1)
comp f g x = f (g x)\\[-2ex]
bench = main [1..20000] []\\[-2ex]
map\_pe0 [] = []
map\_pe0 (x : xs) = ((((x + 1) + 1) + 1) + 1) : map\_pe0 xs
\stopprog
Only two modifications have been performed over the original program:
the annotated expression has been replaced by a call to the new
function \pr{map\_pe0} and the residual (first-order) definition of
\pr{map\_pe0} has been added.  In order to check the improvement
achieved, we can run function \pr{bench} again:
\startprog
map\_iter\_pe> bench
Runtime: 170 msec.
Result: [5,6,7,8,9,10,11,12,13,14,15,16,17,18,19,20,21,22,...]
\stopprog
Thus, the new program runs approximately 4.5 times faster than the
original one. The reason is that it has a first-order definition and
is completely ``deforested'' \cite{Wad90} in contrast to the original
definition. In fact, the most successful experiences were achieved by
specializing calls involving higher-order functions (obtaining
speedups up to a factor of 9) and generic functions with some static
data, like a string pattern matcher where a speedup of 14 was
obtained; experimental results can be found in \cite{AHV02}.

Note that all aforementioned proposals rely on the theoretical
foundations presented in this work. Therefore, our results constitute
the basis for the correctness of all these developments.

\section{Conclusions}\label{sec-concl}

Few attempts have been made to investigate powerful and effective
PE techniques which can be applied to term rewriting systems,
logic programs and functional programs. In this work, we have
introduced the theoretical basis for the PE of 
functional logic programs based
on needed narrowing. We have proved its strong correctness,
i.e., that the answers computed by needed narrowing in the original and
specialized programs for the considered goals are identical (up
to renaming). Furthermore, we have proved that the 
PE process keeps the inductively sequential structure of
programs so that the needed narrowing strategy can also be
used for the execution of specialized programs. As a consequence, our 
PE process preserves the following desirable property for
functional logic programs: deterministic evaluations w.r.t.\ the
original program are still deterministic in the specialized
program. This property is nontrivial as witnessed by
counterexamples for the case of lazy narrowing. 
This allows us to conclude that PE 
based on needed narrowing provides the best known basis
for specializing functional logic programs.

To summarize, the notions presented in this article seem to be the
most promising approach for the PE of modern functional logic
languages based on a lazy semantics:
\begin{itemize}
\item 
We have shown that
a partial evaluator based on lazy narrowing may lead
from orthogonal programs to programs outside this class.  This is
clearly improved by PE based on needed narrowing as it
preserves the original (inductively sequential) structure of
programs, which is the only requirement for the completeness of
the method.

\item On the other hand, modern functional logic languages are
based on (some form of) needed narrowing and, thus, this article is 
intended to be the foundational work in this area.
\end{itemize}
Finally, as we mentioned before, current approaches to the PE of 
multi-paradigm functional logic languages
\cite{AAHV99,AHV02} rely on the theoretical foundations 
presented in this work. Therefore, our results provide
the necessary basis for the correctness of all these subsequent 
developments.




\appendix

\section{Inductive Sequentiality of NN-PE}\label{app-nn-pe-is}

In this section we proof Theorem~\ref{theo-nn-pe-is-terms}
which states that partially evaluated programs are inductively
sequential if the input programs have the same property.
Firstly, this is only proved for PE
w.r.t.\ linear patterns and, then, we extend this result
to arbitrary sets of terms.

\begin{theorem}\label{theo-nn-pe-is}
Let $\Rc$ be an inductively sequential program and $t$ be a linear
pattern. If $\Rc'$ is a pre-NN-PE of $t$ in $\Rc$, then $\Rc'$
is inductively sequential.
\end{theorem}
\begin{proof}
Due to the definition of pre-NN-PE, $\Rc'$ has the form
\[
\begin{array}{l}
\sigma_1(t) \to t_1 \\
\vdots \\
\sigma_n(t) \to t_n
\end{array}
\]
where $t \leadsto_{\sigma_i}^{+} t_i$, $i=1,\ldots,n$, are all
the derivations in the needed narrowing tree for $t$ ending
in a non-failing leaf.
To show the inductive sequentiality of $\Rc'$, it suffices
to show that there exists a definitional tree for the set
$S = \{\sigma_1(t),\ldots,\sigma_n(t)\}$ with pattern $f(\ol{x_p})$
if $t$ has the $p$-ary function $f$ at the root.
We prove this property by induction on the number of inner nodes
of the narrowing tree for $t$.

\noindent
{\bf Base case:} If the number of inner nodes is 1, we first construct
a definitional tree for the set $S = \{t\}$ containing only
the pattern at the root of the narrowing tree.
This is always possible by Proposition~\ref{prop-dt-linear-pattern}.
Now we construct a definitional tree for the sons of the
root by extending this initial definitional tree.
This construction is identical to the induction step.

\noindent
{\bf Induction step:} Assume that $s$ is a leaf in the narrowing
tree,
$\sigma$ is the accumulated substitution from the root to this leaf,
and $\dt$ is a definitional tree for the set
\[
\begin{array}{l}
S = \{ \theta(t) \mid t \leadsto_{\theta}^{+} s'
 \mbox{ is a derivation in the needed narrowing tree}\\
 \mbox{\hspace{20ex} with a non-failing leaf }
s'\}.
\end{array}
\]
Now we extend the narrowing tree by applying one needed narrowing step
to $s$, i.e., let
\[
\begin{array}{l}
s \leadsto_{\phi_1} s_1 \\
\vdots \\
s \leadsto_{\phi_m} s_m
\end{array}
\]
be all needed narrowing steps for $s$.
For the induction step, it is sufficient to show
that there exists a definitional tree for
\[
S' = (S \backslash \{\sigma(t)\}) \cup
 \{\phi_1(\sigma(t)),\ldots,\phi_m(\sigma(t))\}~.
\]
Consider for each needed narrowing step $s \leadsto_{\phi_i} s_i$
the associated canonical representation
$(p,R,\phi_{ik_i} \circ\cdots\circ \phi_{i1}) \in \nns(s,\dt_s)$
(where $\dt_s$ is a definitional tree for the root of $s$). Let
\[
\dt' = \dt \cup \{\phi_{ij}\circ\cdots\circ\phi_{i1}\circ\sigma(t) \mid
1 \leq i \leq m, 1 \leq j \leq k_i \}
\]
We prove that $\dt'$ is a definitional tree for $S'$ by showing that each
of the four properties of a definitional tree holds for $\dt'$.
\begin{description}
\item[\mbox{\rm\em Root property:}]
The minimum elements are identical for both definitional trees, i.e.,
$pattern(\dt) = pattern(\dt')$, since only instances of a leaf of
$\dt$ are added in $\dt'$.

\item[\mbox{\rm\em Leaves property:}]
The set of maximal elements of $\dt$ is $S$. Since all substitutions
computed by needed narrowing along different derivations
are independent
(Lemma \ref{lemma-nn-diff}),
$\sigma$ is independent to all other substitutions occurring in $S$
and the substitutions $\phi_1,\ldots,\phi_m$ are pairwise independent.
Thus, the replacement of the element $\sigma(t)$ in $S$ by the set
$\{\phi_1(\sigma(t)),\ldots,\phi_m(\sigma(t))\}$ does not introduce
any comparable (w.r.t.\ the subsumption ordering) terms.
This implies that $S'$ is the set of maximal elements of $\dt'$.

\item[\mbox{\rm\em Parent property:}]
Let $\pi \in \dt' \backslash \{pattern(\dt')\}$. We consider two cases
for $\pi$:
\begin{enumerate}
\item $\pi \in \dt$:
Then the parent property trivially holds since
only instances of a leaf of $\dt$ are added in $\dt'$.
\item $\pi \not\in \dt$:
By definition of $\dt'$,
$\pi = \phi_{ij}\circ\cdots\circ\phi_{i1}\circ\sigma(t)$ for some
$1 \leq i \leq m$ and $1 \leq j \leq k_i$. We show by induction on $j$
that the parent property holds for $\pi$.\\
Base case ($j=1$): Then $\pi = \phi_{i1}(\sigma(t))$.
It is $\phi_{i1} \neq id$ (otherwise $\pi = \sigma(t) \in \dt$).
Thus, by Proposition~\ref{prop-nn-subst},
$\phi_{i1} = \{x \mapsto c(\ol{x_n})\}$
with $x \in \Var(s) \subseteq \Var(\sigma(t))$.
Due to the linearity of the initial pattern and all substituted terms
(cf.\ Proposition~\ref{prop-nn-subst}),
$\sigma(t)$ has a single occurrence $o$ of the variable $x$ and,
therefore, $\pi = \sigma(t)[c(\ol{x_n})]_o$, i.e.,
$\sigma(t)$ is the unique parent of $\pi$.\\
Induction step ($j>1$): We assume that the parent property holds for
$\pi' = \phi_{i,j-1}\circ\cdots\circ\phi_{i1}\circ\sigma(t)$.
Let $\phi_{ij} \neq id$ (otherwise the induction step is trivial).
By Proposition~\ref{prop-nn-subst}, $\phi_{ij} = \{x \mapsto c(\ol{x_n})\}$
with $x \in \Var(\phi_{i,j-1}\circ\cdots\circ\phi_{i1}\circ\sigma(t))$
(since $\Var(s) \subseteq \Var(\sigma(t))$).
Now we proceed as in the base case to show that
$\pi'$ is the unique parent of $\pi$.
\end{enumerate}

\item[\mbox{\rm\em Induction property:}]
Let $\pi \in \dt' \backslash S'$. We consider two cases for $\pi$:
\begin{enumerate}
\item $\pi \in \dt \backslash \{\sigma(t)\}$:
Then the induction property holds for $\pi$ since it already holds in $\dt$
and only instances of $\sigma(t)$ are added in $\dt'$.
\item $\pi = \phi_{ij}\circ\cdots\circ\phi_{i1}\circ\sigma(t)$ for some
$1 \leq i \leq m$ and $0 \leq j < k_i$.
Assume $\phi_{i,j+1} \neq id$ (otherwise, do the identical proof
with the representation
$\pi = \phi_{i,j+1}\circ\cdots\circ\phi_{i1}\circ\sigma(t)$).
By Proposition~\ref{prop-nn-subst},
$\phi_{i,j+1} = \{x \mapsto c(\ol{x_n})\}$ and $\pi$ has
a single occurrence of the variable $x$ (due to the linearity of the
initial pattern and all substituted terms). Therefore,
$\pi' = \phi_{i,j+1}\circ\cdots\circ\phi_{i1}\circ\sigma(t)$
is a child of $\pi$. Consider another child
$\pi'' = \phi_{i'j'}\circ\cdots\circ\phi_{i'1}\circ\sigma(t)$
of $\pi$ (other patterns in $\dt'$ cannot be children of $\pi$
due to the induction property for $\dt$). Assume
$\phi_{i'j'}\circ\cdots\circ\phi_{i'1} \neq
\phi_{i,j+1}\circ\cdots\circ\phi_{i1}$
(otherwise, both children are identical).
By Lemma~\ref{lemma-nn-diff}, there exists some $l$ with
$\phi_{i'l}\circ\cdots\circ\phi_{i'1} = \phi_{il}\circ\cdots\circ\phi_{i1}$,
$\phi_{i',l+1} = \{x' \mapsto c'(\cdots)\}$, and
$\phi_{i,l+1} = \{x' \mapsto c''(\cdots)\}$ with $c' \neq c''$.
Since $\pi''$ and $\pi'$ are children
of $\pi$ (i.e., immediate successors w.r.t.\ the subsumption ordering),
it must be $x' = x$ (otherwise, $\pi'$ differs from $\pi$ at more than
one position) and $\phi_{i',j'}= \cdots = \phi_{i',l+2} = id$
(otherwise, $\pi''$ differs from $\pi$ at more than one position).
Thus, $\pi'$ and $\pi''$ differ only in the instantiation
of the variable $x$ which has exactly one occurrence in their
common parent $\pi$, i.e., there is a position
$o$ of $\pi$ with $\pi|_o = x$ and $\pi' = \pi[c'(\ol{x_{n'_i}})]_o$
and $\pi'' = \pi[c''(\ol{x_{n''_i}})]_o$.
Since $\pi''$ was an arbitrary child of $\pi$, the induction
property holds.
\end{enumerate}
\end{description}
\end{proof}
Since actual partial evaluations are usually computed for more than one term,
we extend the previous theorem to this more general case.

\begin{corollary}\label{cor-nn-pe-is}
Let $\Rc$ be an inductively sequential program and $S$ be a finite set
of linear patterns with pairwise different root symbols.
If $\Rc'$ is a pre-NN-PE of $S$ in $\Rc$, then $\Rc'$
is inductively sequential.
\end{corollary}
\begin{proof}
This is a consequence of Theorem~\ref{theo-nn-pe-is}
since we can construct a definitional tree
for each pre-NN-PE of a pattern in $S$.
Since all patterns have different
root symbols, the roots of these definitional trees
do not overlap.
\end{proof}
Now we are able to show that the NN-PE of an arbitrary
set of terms---w.r.t.\ an inductively sequential program---always produces
an inductively sequential program. \\

\noindent
{\it Theorem~\ref{theo-nn-pe-is-terms}}\\
Let $\Rc$ be an inductively sequential program and $S$ a finite
set of operation-rooted terms. Then each NN-PE of $\Rc$ w.r.t.\ $S$
is inductively sequential.

\begin{proof}
Let $\Rc'$ be a pre-NN-PE of $\Rc$ w.r.t.\ $S$
and $\rho$ an independent renaming of $S$.
Then each rule of a NN-PE $\Rc''$ of $\Rc$ w.r.t.\ $S$ (under $\rho$)
has the form $\theta(\rho(s)) \to ren_{\rho}(r)$ for some rule
$\theta(s) \to r \in \Rc'$. Consider the extended
rewrite system
\[
\Rc_\rho = \Rc \cup \{\rho(s) \to s \mid s \in S \}
\]
where the renaming $\rho$ is encoded by a set of rewrite rules.
Note that $\Rc_\rho$ is inductively sequential since the new
left-hand sides $\rho(s)$ are of the form $f_s(\ol{x_n})$
with new function symbols $f_s$.

Let $\Rc'_\rho$ be an arbitrary pre-NN-PE of $\Rc_\rho$ w.r.t.\ $\rho(S)$.
Since $\rho(S)$ is a set of linear patterns with pairwise different
root symbols, $\Rc'_\rho$ is inductively sequential by
Corollary~\ref{cor-nn-pe-is}.
It is obvious that each subset of an inductively sequential
program is also inductively sequential (since only the left-hand sides
of the rules are relevant for this property).
Therefore, to complete the proof it is sufficient to show
that all left-hand sides of rules from $\Rc''$ can also
occur as left-hand sides in some $\Rc'_\rho$.

Each rule of $\Rc''$ has the form $\theta(\rho(s)) \to ren_{\rho}(r)$
for some rule $\theta(s) \to r \in \Rc'$.
By definition of $\Rc'$, there exists a needed narrowing derivation
$s \leadsto_\theta^{+} r$ w.r.t.\ $\Rc$.
Hence,
\[
 \rho(s) ~\leadsto_{id}~ s ~\leadsto_\theta^{+}~ r
\]
is a needed narrowing derivation w.r.t.\ $\Rc_\rho$.
Thus, $\theta(\rho(s)) \to r$ is a resultant which can occur
in some $\Rc'_\rho$.
\end{proof}

\section{Strong Correctness of NN-PE}\label{sc-nn-pe}

In this section, we prove Theorem~\ref{theo-nnpe-strong-correct},
i.e., the strong correctness of NN-PE,
and introduce some necessary auxiliary notions and results
for this proof.
The proof proceeds essentially as follows. Firstly, we prove the soundness
(resp.\ completeness) of the transformation, i.e., we prove that for
each answer computed by needed narrowing
in the original (resp.\ specialized) program there
exists a more general answer in the specialized (resp.\ original) program
for the considered queries.
Then, by using the minimality of needed narrowing,
we conclude the strong correctness of NN-PE,
i.e., the answers computed in the original and the partially evaluated
programs coincide (up to renaming).

In order to simplify the proofs, we assume (without loss of generality)
that the rules of strict equality are automatically
added to the original as well as to the partially evaluated
program. We also assume that the set of specialized
terms always contains the calls $x \equ y$ and
$x \wedge y$, and by abuse we consider that $\rho$ does not
modify these symbols.
This allows us to handle the strict equality rules in $\cR'$ as ordinary
resultants derived from the one-step needed
narrowing derivations for the calls  $x \equ y$ and $x \wedge y$
in $\cR$.

\subsection{Soundness}

The following lemmata are auxiliary to prove that reduction sequences in
the specialized program can also be performed in the original program
(up to renaming of terms and programs).

\begin{lemma} \label{resultant}
Let $\cR$ be an inductively sequential program and $s$ be an
operation-rooted term.
Let $s \leadsto_{\sigma}^{+} r$ be a needed narrowing
derivation w.r.t.\ $\cR$ whose associated resultant is
$R = (\sigma(s) \rightarrow r)$.
If $t \rightarrow_{p,R} t'$ for some position $p \in \pos(t)$, then
$t \rightarrow^{+} t'$ w.r.t.\ $\cR$.
\end{lemma}

\begin{proof}
Given the  derivation $s \leadsto_{\sigma}^{+} r$,
by soundness of needed narrowing (claim 1 of
Theorem~\ref{theo-needed-properties}), we have
$\sigma(s) \rightarrow^{+} r$. Since $t \rightarrow_{p,R} t'$,
there exists a substitution $\theta$ such that
$\theta(\sigma(s)) = t|_{p}$ and $t' = t[\theta(r)]_{p}$.
Since $\sigma(s) \rightarrow^{+} r$,
by stability of rewriting,
we have $\theta(\sigma(s)) \rightarrow^{+} \theta(r)$.
Therefore $t = t[\theta(\sigma(s))]_p
\rightarrow^{+} t[\theta(r)]_p = t'$ w.r.t.\ $\cR$,
which concludes the proof.
\end{proof}

\begin{lemma} \label{sound-aux}
Let $S$ be a finite set of terms and $\rho$ an independent
renaming for $S$.
Let $R = (\theta(s) \rightarrow r)$ be a rewrite rule such that
$\theta$ is constructor and $s \in S$, and let
$R' = (l' \rightarrow r')$ be a renaming of $R$ where
$l' = \theta(\rho(s))$ and $r' = ren_{\rho}(r)$. Given a term
$t_1$ and one of its renamings
$t'_1 = ren_{\rho}(t_1)$, if $t'_1 \rightarrow_{p',R'} t'_2$
then $t_1 \rightarrow_{p,R} t_2$ where $p$ is the corresponding position
of $p'$ in $t'_1$ and $t'_2 = ren_{\rho}(t_2)$.
\end{lemma}
\begin{proof}
Immediate by definition of $ren_{\rho}$.
\end{proof}

The following proposition is the key to prove the soundness of NN-PE.

\begin{proposition} \label{main-sound}
Let $\cR$ be an inductively sequential program. Let $e$ be an equation,
$S$ a finite set of operation-rooted
terms, $\rho$ an independent renaming of $S$,
and $\cR'$ a NN-PE of $\cR$ w.r.t.\ $S$ (under $\rho$) such that
$\cR' \cup \{e'\}$
is $S'$-closed, where $e' = ren_{\rho}(e)$ and $S' = \rho(S)$.
If $e' \rightarrow^{\ast} true$ in $\cR'$ then
$e \rightarrow^{\ast} true$ in $\cR$.
\end{proposition}
\begin{proof}
We prove the claim by induction on the number $n$ of rewrite
steps in $e' \rightarrow^{\ast} true$ (considering $e'$ an
arbitrary $S'$-closed expression).
\begin{description}
\item[\rm Base case.] If $n = 0$, we have $e' = true$
and the claim trivially follows since $ren_{\rho}(true) = true$
by definition.

\item[\rm Inductive case.]
Consider a rewrite sequence of the form
$e' \rightarrow_{p',R'} h' \rightarrow^{\ast} true$ with
$R' = (l' \rightarrow r')$.
By definition of NN-PE, $R'$ has been obtained by applying the
post-processing renaming to a rule $R = (\theta(s) \rightarrow r)$
in the pre-NN-PE, where $\theta$ is constructor,
$l' = \theta(\rho(s))$, and $r' = ren_{\rho}(r)$.
By Lemma~\ref{sound-aux}, we have $e \rightarrow_{p,R} h$
where $p$ is the corresponding position of $p'$ in $e'$
and $h' = ren_{\rho}(h)$.
By definition of pre-NN-PE, there exists
a needed narrowing derivation
$s \leadsto^{+}_{\theta} r$ which produced the resultant $R$.
Since $e \rightarrow_{p,R} h$, we have $e \rightarrow^{+} h$
in $\cR$ by Lemma~\ref{resultant}.

Since the terms in $S'$ are linear and $\cR'$ is
$S'$-closed, $h'$ is trivially $S'$-closed.
By applying the inductive hypothesis to the subderivation
$h' \rightarrow^\ast true$ in $\cR'$,
there exists a sequence $h \rightarrow^\ast true$ in $\cR$. Together
with the initial sequence $e \rightarrow^{\ast} h$  we
get the desired derivation in $\cR$.
\end{description}
\end{proof}
Now we state and prove the soundness of NN-PE.

\begin{theorem} \label{wsoundness}
Let $\cR$ be an inductively sequential program. Let $e$ be an equation,
$V \supseteq \var(e)$ a finite set of variables, $S$ a
finite set of operation-rooted terms, and $\rho$ an
independent renaming of $S$. Let $\cR'$ be a
NN-PE of $\cR$ w.r.t.\ $S$ (under $\rho$)
such that $\cR' \cup \{e'\}$ is $S'$-closed,
where $e' = ren_{\rho}(e)$ and $S' = \rho(S)$. If $e'
\leadsto_{\sigma'}^{\ast} true$ is a needed narrowing
derivation for $e'$ in $\cR'$,
then there exists a needed
narrowing derivation $e \leadsto_{\sigma}^{\ast} true$ in $\cR$ such
that $(\sigma \leq \sigma')~[V]$.
\end{theorem}

\begin{proof}
Since $e' \leadsto^{\ast}_{\sigma'} true$ in $\cR'$ and $\cR'$ is
inductively sequential (Theorem~\ref{theo-nn-pe-is-terms}), by the soundness
of needed narrowing (claim 1 of Theorem~\ref{theo-needed-properties}),
we have $\sigma'(e') \rightarrow^{\ast} true$.
Since $e'$ is $S'$-closed and $\sigma'$
is constructor, by definition of closedness, $\sigma'(e')$ is also
$S'$-closed and $\sigma'(e') = ren_{\rho}(\sigma'(e))$.
By Proposition~\ref{main-sound},
there exists a rewrite sequence $\sigma'(e) \rightarrow^{\ast} true$ in
$\cR$. Therefore, by the completeness of needed narrowing
(claim 2 of Theorem~\ref{theo-needed-properties}),
there exists a needed narrowing derivation
$e \leadsto_{\sigma}^{\ast} true$ in $\cR$ such that $(\sigma \leq
\sigma') ~[V]$, which completes the proof.
\end{proof}

\subsection{Completeness}

Firstly, we consider the notions of descendants and traces.
Let $A = (t \rightarrow_{u,l \rightarrow r} t')$ be a reduction step
of some term $t$ into $t'$ at position $u$ with rule $l \rightarrow r$.
The set of \emph{descendants} \cite{HL92}
of a position $v$ of $t$ by $A$, denoted $v \backslash A$, is
\[ v \backslash A = \left\{
\begin{array}{ll}
\emptyset & \mbox{if } u = v,  \\
\{ v \} & \mbox{if } u \not\leq v, \\
\{ u.p'.q \mid r|_{p'} = x \} & \mbox{if }
  v = u.p.q \mbox{ and } l|_p = x, \mbox{ where } x \in \cX.
\end{array} \right.
\]
The set of \emph{traces}
of a position $v$ of $t$ by $A$, denoted $v \traces A$ is
\[ v \traces A = \left\{
\begin{array}{ll}
\{ v \} & \mbox{if } u = v,  \\
\{ v \} & \mbox{if } u \not\leq v, \\
\{ u.p'.q \mid r|_{p'} = x \} & \mbox{if }
  v = u.p.q \mbox{ and } l|_p = x, \mbox{ where } x \in \cX.
\end{array} \right.
\]
The set of descendants of a position $v$ by a reduction sequence $B$
is defined inductively as follows
\[ v \backslash B = \left\{
\begin{array}{ll}
\{ v \} & \mbox{if $B$ is the null derivation,} \\
\displaystyle \bigcup_{w \in v \backslash B'} w \backslash B''
  & \mbox{if $B = B'B''$, where $B'$ is the initial step of $B$.}
\end{array} \right.
\]
Given a set of positions $P$, we let $P \backslash B =
\bigcup_{p \in P} ~ p \backslash B.$ The definition of the set of traces
of a position by a reduction sequence is perfectly analogous.

A redex $s$ in a term $t$ is root-needed, if $s$ (itself or one of
its descendants) is contracted in every rewrite sequence from $t$ to a
root-stable term \cite{Mid97}.

In the remainder of this section, we consider \emph{outermost-needed}
reduction sequences as defined\footnote{This is a slightly
different though equivalent definition, since
we do not allow for \emph{exempt} nodes, as in
\cite{Ant92}.} in \cite{AEH00}.

\begin{definition}[\citeNP{Ant92}]\label{DefOutNeedStrat}
Let $\Rc$ be an inductively sequential program.
The (partial) function $\varphi$ takes arguments $t=f(\ol{t})$ for
a given
$f\in{\cal F}$, and a definitional tree\footnote{In this definition,
we write $branch(\pi,p,\dt_1,\ldots,\dt_n)$ for a definitional tree
$\dt$ with pattern $\pi$ if $\pi$ is a branch node with inductive
position $p$ and children $\pi_1,\ldots,\pi_n$ where
$\dt_i = \{\pi' \in \dt \mid \pi_i \leq \pi'\}$, $i=1,\ldots,n$.} $\dt$
such that $pattern(\dt)\leq t$, and yields a
redex occurrence $p\in \pos_{\cal R}(t)$ called an
\emph{outermost-needed} redex:
\[ \varphi(t,\dt)~=
\left \{
\begin{array}{l@{~~~}ll}
\toppos & \mbox{if $\dt=\{\pi\}$}\\
\varphi(t,\dt_i) & \mbox{if $\dt=branch(\pi,p,\dt_1,\ldots,\dt_n)$}\\
& ~~~\mbox{and $pattern(\dt_i)\leq t$ for some $i$, $1 \leq i \leq n$}\\
p.\varphi(t|_p,\dt_g) & \mbox{if
$\dt=branch(\pi,p,\dt_1,\ldots,\dt_n)$,}\\
& ~~~\mbox{$root(t|_p)=g\in{\cal F}$, and}\\
& ~~~\mbox{$\dt_g$ is a definitional tree for $g$}.
\end{array}
\right.\]
\end{definition}
The following notations and terminology are needed for the subsequent 
developments. Positions $u,v$ are \emph{disjoint}, 
denoted $u \:\bot\: v$, if neither $u\leq v$ nor $v \leq u$.
Also, for a set of (pairwise disjoint, ordered) positions 
$P=\{p_{1},\ldots,p_{n}\}$, we let
$t[s_{1},\ldots,s_{n}]_P =
(((t[s_{1}]_{p_{1}})[s_{2}]_{p_{2}})\ldots[s_{n}]_{p_{n}})$.
A term is root-normalizing if it has a root-stable reduct.
If there are rules $l \to r$ and $l' \to r'$ and a
most general unifier $\sigma$ for $l|_p$ and $l'$ for some position $p$,
the pair $\langle\sigma(l)[\sigma(r')]_p,\sigma(r)\rangle$ is called a
\emph{critical pair} and it is also called an \emph{overlay}
if  $p=\toppos$. A  critical pair $\langle t,s\rangle$
is trivial if $t=s$. A TRS is  called
\emph{almost orthogonal} if its critical pairs are trivial overlays. 
If all critical pairs are trivial, a TRS 
is  called \emph{weakly orthogonal}. Note that,
in CB-TRSs, almost orthogonality and weak orthogonality coincide.
The inner reduction relation is
$\inr = \mathord{\to} \backslash \exr$.
The following technical results are auxiliary.

\begin{lemma}[\citeNP{Mid97}] \label{LemaMiddeldorp3_3}
Let ${\cal R}$ be an almost orthogonal TRS. If $t$ is root-stable and
$s\inrs t$, then $s$ is root-stable.
\end{lemma}

\begin{theorem}
\label{TheoOutNeedIndAreRootNeeded}
Let ${\cal R}$ be an inductively sequential program and
$t$ be a non-root-stable term. Every outermost-needed redex is
root-needed.
\end{theorem}

\begin{proof}
By Theorem 18 in \cite{HanLucMid98},
outermost-needed redexes are addressed by strong indices. By Theorem
5.6 in \cite{Luc98}, nv-indices (hence strong
indices, see \cite{Oya93}) in
non-root-stable terms address root-needed redexes.
\end{proof}

\begin{theorem}\label{TheoSubTermAndRed}
Let $\cR$ be a weakly orthogonal CB-TRS and $t$ be a term. Let
$P=\{p_1,\ldots,p_n\}\subseteq\pos(t)$ be a set of disjoint positions
of $t$ such that each $t|_{p_i}$ for $1\leq i\leq n$ is operation-rooted.
If $t$ admits a root-normalizing derivation which does not root-normalize any
$t|_{p_i}$, then $t$ admits a root-normalizing derivation which does not
reduce any $t|_{p_i}$.
\end{theorem}

\begin{proof}
If $t$ is root-stable, the result is immediate. If $t$ is not
root-stable, then there exists a root-stable reduct $\sigma(r)$
and a derivation
$A:t\to^\ast\sigma(l)\:\exr\:\sigma(r)$ for some rule
$l\to r$ in $\cR$ which, by hypothesis, root-normalizes $t$ without
root-normalizing any
$t|_{p_i}$. Let $\ol{y_n}=y_1,\ldots,y_n$ be new, distinct variables each of which is used to
name a subterm $t|_{p_i}$. The substitution
$\theta_t$ defined by $\theta_t(y_i)=t|_{p_i}$ associates a subterm to each variable.
Note that $\theta_t(t[\ol{y_n}]_P)=t$.
As an intermediate step of the demonstration, first we prove,
by induction on the length $N+1$ of the derivation
$A$, that there exists a substitution $\sigma'$ such that $t[\ol{y_n}]_P\to^\ast\sigma'(l)$
and $\theta_t(\sigma'(x))\to^\ast\sigma(x)$ for all $x\in\Var(l)$.

First we note that, since the derivation $A$ does not
root-normalize any $t|_{p_i}$, we have that $p_i>\Lambda$ for every
$1\leq i\leq n$ (otherwise $P=\{\Lambda\}$ and we obtain a
contradiction with the initial hypothesis).
\begin{enumerate}
\item If $N=0$, then $t=\sigma(l)$. Since $p_i>\Lambda$, $t|_{p_i}$ is
operation-rooted for $1\leq i\leq  n$, and $\cR$ is constructor
based, then for each $p_i$ there exists a variable position $v_i\in \pos(l)$
such that $p_i=v_i.w_i$
and $l|_{v_i}$ is a variable.
Then, for each $x\in\Var(l)$, we let $\sigma'(x)=t[\ol{y_{n}}]_P|_{v_x}$ where
$v_x$ is the position of $x$ in $l$.
Hence, $t[\ol{y_n}]_P=\sigma'(l)$ and
$\theta_t(\sigma'(x))=\theta_t(t[\ol{y_{n}}]_P|_{v_x})=\sigma(x)$ for
each $x\in\Var(l)$.
Thus, $\theta_t(\sigma'(x))\to^\ast\sigma(x)$.
\item If $N>0$, then we consider the derivation
$t\to_{q}t'\to^\ast\sigma(l)$. Let $P'=q\traces
P=\{p'_1,\ldots,p'_{n'}\}$ be the
traces of $P$ w.r.t.\ the rewriting step $t\to_{q}t'$
(note that the traces are well defined since every $t|_{p_i}$
is operation-rooted).
By hypothesis, the derivation
$t\to^\ast\sigma(l)\to \sigma(r)$ does not
root-normalize any $t|_{p}$ for $p\in P$. In particular, the step
$t\to_{q}t'$ does not root-normalize any $t|_p$ for $p\in P$.
Therefore, each $t'|_{p'}$ for every
$p'\in P'$ is operation-rooted and the derivation
$t'\to^\ast\sigma(l)\to \sigma(r)$ does not
root-normalize any $t'|_{p'}$ for $p'\in P'$. Thus, by the induction
hypothesis, $t'[\ol{z_{n'}}]_{P'}\to^\ast\sigma'(l)$  and
$\theta_{t'}(\sigma'(x))\to^\ast\sigma(x)$ for all
$x\in\Var(l)$ where $\ol{z_{n'}}=z_1,\ldots,z_{n'}$ are new, distinct
variables which identify the subterms in $t'$ addressed by $P'$, i.e.,
$\theta_{t'}(z_i)=t'|_{p'_i}$ for $1\leq i\leq n'$. We connect variables in
$\ol{z_{n'}}$ and variables in $\ol{y_{n}}$ by means of a substitution
$\tau:\ol{z_{n'}}\to \ol{y_{n}}$ as follows: $\tau(z_i)=y_j$ iff $p'_i$ is a
trace of $p_j$ (w.r.t.\
the step $t\to t'$) for
$1\leq i\leq n'$ and $1\leq j\leq n$.
Now we consider two cases:
\begin{enumerate}
\item If there is no $p\in P$ such that $p\leq q$, then, since each
$t|_{p_i}$ is operation-rooted and $\cR$ is constructor-based, we
have that $t[\ol{y_{n}}]_{P}\to_{q}
\tau(t'[\ol{z_{n'}}]_{P'})$. Moreover,
since no $t|_{p_i}$ changes in this rewriting step, we have
$\theta_t(\tau(z))=\theta_{t'}(z)$ for all $z\in \ol{z_{n'}}$, i.e.,
$\theta_{t'}=\theta_t\circ\tau$. Since
$t'[\ol{z_{n'}}]_{P'}\to^\ast\sigma'(l)$, by stability,
$\tau(t'[\ol{z_{n'}}]_{P'})\to^\ast\tau(\sigma'(l))$. Thus,
$t[\ol{y_{n}}]_{P}\to^\ast\tau(\sigma'(l))$. Since
$\theta_{t'}(\sigma'(x))=\theta_t(\tau(\sigma'(x)))\to^\ast\sigma(x)$,
 the conclusion follows.
\item If there is $p\in P$ such that $p\leq q$, then $P'=P$, $n=n'$
and we can take $\ol{y_n}=\ol{z_{n'}}$. Hence,
$t[\ol{y_{n}}]_P=t'[\ol{y_{n}}]_P=t'[\ol{z_{n'}}]_{P'}$. Now we have that
$\theta_{t}(y_i)\to\theta_{t'}(y_i)$ if $p=p_i$, for some $1\leq
i\leq n$ whereas
$\theta_{t}(y_j)=\theta_{t'}(y_j)$ for all $j\neq i$, and the conclusion also
follows.
\end{enumerate}
\end{enumerate}
Since $\theta_t(\sigma'(x))\to^\ast\sigma(x)$ for all
$x\in\Var(l)$, we consider two possibilities:
\begin{enumerate}
\item If $r\not\in\cX$, then, since $\Var(r)\subseteq\Var(l)$, we
have that $\theta_t(\sigma'(r))\inrs\sigma(r)$.
\item If $r=x\in\cX$, we prove that this implies that
$\theta_t(\sigma'(r))=\theta_t(\sigma'(x))\inrs$
$\sigma(x)=\sigma(r)$.
Otherwise, it is necessary that $\sigma'(x)$ be a variable. In
this case, it must be $\sigma'(x)=y_i$ for some $1\leq i\leq n$
(otherwise, $\theta_t(\sigma'(x))=\sigma'(x)$ is a variable and it
cannot be rewritten to $\sigma(x)$ in zero or more steps unless
$\sigma'(x)=\sigma(x)$ in
which case, we trivially have that $\theta_t(\sigma'(x))\inrs\sigma(x)$).
Since $\sigma(r)=\sigma(x)$ is root-stable, the existence of the
derivation $\theta_t(\sigma'(x))=\theta_t(y_i)\to^\ast\sigma(x)$ implies (since each
reduction step in the derivation $\theta_t(\sigma'(x))\to^\ast\sigma(x)$
has been taken from the derivation $A$) that the
derivation $A$ root-normalizes the subterm $\theta_t(y_i)=t|_{p_i}$.
This contradicts our initial hypothesis.
\end{enumerate}
Thus, in all cases, we have that $\theta_t(\sigma'(r))\inrs\sigma(r)$ and, since
$\sigma(r)$ is root-stable, by Lemma \ref{LemaMiddeldorp3_3}
(remember that weak orthogonality and almost orthogonality
coincide for CB-TRSs),
$\theta_t(\sigma'(r))$ is root-stable. Note that we have also proved
that $t[\ol{y_{n}}]_P\to^\ast\sigma'(l)\to\sigma'(r)$ and therefore, by stability,
$t=\theta_t(t[\ol{y_{n}}]_P)\to^\ast\theta_t(\sigma'(r))$ is a root-normalizing
derivation for $t$ which does not reduce any $t|_{p_i}$ for $1\leq i\leq n$.
\end{proof}

\begin{theorem}\label{TheoRootStabSubtAndRed}
Let $\cR$ be a weakly orthogonal CB-TRS and $t$ be a term. Let
$P=\{p_1,\ldots,p_n\}\subseteq\pos(t)$ be a set of disjoint positions
of $t$ such that each $t|_{p_i}$ for $1\leq i\leq n$ is a
root-stable, operation-rooted term.
If $t$ is root-normalizing, then $t$ admits a root-normalizing
derivation which does not reduce any $t|_{p_i}$.
\end{theorem}

\begin{proof}
The proof is perfectly analogous to the proof of Theorem
\ref{TheoSubTermAndRed}.
Assume the same notations for the proof.
Only one difference arises in the last part of the proof: we need not
distinguish  the cases $r\in\cX$ and $r\not\in\cX$. This is because
the fact that each $t|_{p_i}$ is root-stable and the fact that
$\theta_t(\sigma'(x))\to^\ast\sigma(x)$ for all $x\in\Var(l)$ allows us
to immediately derive that $\theta_t(\sigma'(x))\inrs\sigma(x)$. Now, it
suffices to consider that every $t|_{p_i}$ and their possible reducts are
operation-rooted, which easily follows from the fact that each  $t|_{p_i}$ is
root-stable and operation-rooted.
\end{proof}

\begin{theorem}\label{TheoNoRootNeededRedex}
Let $\cR$ be a weakly orthogonal CB-TRS, $t$ be a term, and $p\in\pos(t)$. Let
$s$ be a root-stable, operation-rooted subterm of $t$.
If $t$ is root-normalizing, then $s$ does not have redexes which are
root-needed in $t$.
\end{theorem}

\begin{proof}
Immediate, by using Theorem \ref{TheoRootStabSubtAndRed}.
\end{proof}

\begin{theorem}\label{TheoRootNeededInSubtermAndRootNeededTerm}
Let $\cR$ be a weakly orthogonal CB-TRS and $t$ be a term. If $s$ is an
operation-rooted subterm of $t$
that contains a redex which is root-needed in $t$, then every
root-needed redex in $s$ is root-needed in $t$.
\end{theorem}

\begin{proof}
Since $t$ contains at least a root-needed redex, then $t$ is not
root-stable. If $t$ has no root-stable form, it is trivial, since every redex is
root-needed in $t$. Hence, we assume that $t$ has a root-stable reduct.
Let $s|_q$ be a root-needed redex in $s$.
Then, $s$ is not root-stable.
Let $s|_{q'}$ be a root-needed redex in $t$.
If $s|_{q}$ is not root-needed in $t$,
then it is possible to root-normalize $t$ without reducing the redex
$s|_{q}$.
However, without reducing the redex
$s|_q$ it is not possible to root-normalize $s$. Therefore,
it is possible to root-normalize $t$ without root-normalize $s$.
By Theorem \ref{TheoSubTermAndRed}, it is possible to root-normalize
$t$ without reducing $s$, hence without reducing $s|_{q'}$, which
yields a contradiction.
\end{proof}

\noindent
The following auxiliary definition is useful to deal with closed terms
(it is a slight refinement of the same notion in \cite{AFV98}).

\begin{definition}[covering set, closure set] \label{cset}
Let $S$ be a finite set of terms and $t$ be an $S$-closed term. We
define the covering set of $t$ w.r.t.\ $S$ as follows:
\[ CSet(S,t) = \{ O \mid O \in c\_set(S,t), ~  (u.0,fail) \not\in O,
~ u \in \nat^\ast \} \]
where the auxiliary function $c\_set$, used to compute each closure
set $O$, is defined inductively as follows:
\[
\begin{array}{@{\hspace{-5ex}}l}
c\_set(S,t) \ni \\
\left\{ \begin{array}{ll}
\emptyset & \mbox{if } t \in \cX \cup \cC, \\
\bigcup_{i=1}^{n}
\{ (i.p,s) \mid (p,s) \in c\_set(S,t_{i}) \}
  & \mbox{if } t = c(\ol{t_{n}}),~ c \in \cC^\ast, \\
\{ (\toppos,s) \} \cup \{ (q.p,s') \mid s|_{q} \in \cX, ~  (p,s') \in
c\_set(S,\theta(s|_{q})) \}
  & \mbox{if }
\exists s \in S \:\mbox{s.t.}\: \theta(s) = t \\
& \mbox{for some } \theta, \\
\{ (0,fail) \} & \mbox{otherwise.}
\end{array}
\right.
\end{array}
\]
where $\cC^\ast = \cC \cup \{\equ,\wedge\}$.
Note that positions ending with the mark ``$0$'' identify the
situation in which some subexpression of $t$ is not an instance of any
of the terms in $S$. Thus, a set containing a pair of the form
$(u.0,fail)$ is not considered a closure set.
\end{definition}

Roughly speaking, given a set of terms $S$ and a term $t$ which is
$S$-closed, each set in $CSet(S,t)$ identifies a concrete way in
which $t$ can be proved $S$-closed, thus avoiding the non-determinism
which is implicit in the definition of closedness.

The following \emph{lifting} lemma is a slight variant of the
completeness result for needed narrowing.

\begin{lemma} \label{lifting}
Let $\cR$ be an inductively sequential program. Let
$\sigma$ be a constructor substitution, $V$ a finite set
of variables, and $s$ an operation-rooted term with $\var(s)
\subseteq V$. If
$\sigma(s) \rightarrow_{p_{1},R_{1}} \cdots
\rightarrow_{p_{n},R_{n}} t$
is an outermost-needed reduction sequence,
then there exists a needed narrowing derivation
$s \leadsto_{p_{1},R_{1},\sigma_{1}} \cdots
\leadsto_{p_{n},R_{n},\sigma_{n}} t'$
and a constructor substitution $\sigma'$ such that
$\sigma'(t') = t$ and $(\sigma' \circ \sigma_{n}
\circ \cdots \circ \sigma_{1} = \sigma)~[V]$.
\end{lemma}

\begin{proof}
It is perfectly analogous to the proof of Theorem 4 (completeness) in
\cite{AEH00}.
\end{proof}
Now we prove two technical results which are necessary for a useful
generalization of
the lifting lemma.
We need to  make the lemma applicable
even when the
considered substitution  is not constructor, as long as it still does
not introduce a needed redex.
In order to do this  extension, we need to ensure that it
is possible to get rid of some
operation-rooted subterms which are introduced by instantiation
whenever they  are not contracted in
the considered derivation. We prove this in the following lemmata.

\begin{lemma} \label{liftaux}
Let $\cR$ be program. Let $t$ and $s$ be operation-rooted
terms and $P_0 \subseteq \pos(t)$ be a nonempty set of
disjoint positions such that $t|_{p} = s$ for all $p \in P_0$.
Let
\[ t[s,\ldots,s]_{P_0} = t_0 \rightarrow_{p_{1},R_{1}} \cdots
\rightarrow_{p_{n},R_{n}} t_n = t'[s,\ldots,s]_{P_n} \]
be a reduction sequence
where $A_i = (t_{i-1} \rightarrow_{p_{i},R_{i}} t_i)$ and
$P_i = P_{i-1} \backslash A_i$ for all $i = 1,\ldots,n$, $n \geq 0$.
If $p \not\leq p_i$ for all $p \in P_{i-1}$, $i = 1,\ldots,n$,
then there exists a reduction sequence
\[ t[x,\ldots,x]_{P_0} \rightarrow_{p_{1},R_{1}} \cdots
\rightarrow_{p_{n},R_{n}} t'[x,\ldots,x]_{P_n} \]
\end{lemma}

\begin{proof}
By induction on the number $n$ of steps in the former reduction:
\begin{description}
\item[$n=0$.] Trivial.

\item[$n > 0$.] Consider $A_1 = (t[s,\ldots,s]_{P_0}
\rightarrow_{p_{1},R_{1}} t''[s,\ldots,s]_{P_1})$, where
$R_{1} = (l_{1} \rightarrow r_{1})$,
$\sigma_{1}(l_{1}) = t|_{p_{1}}$, and $P_1 = P_0 \backslash A_1$.
We distinguish two cases depending on the relative
position of
$p_{1}$ (the case $p \leq p_{1}$, for some $p \in P_0$, is not considered
since the subterms in $s$ are not contracted, i.e.,
$p \not\leq p_1$ for all $p \in P_0$):
\begin{description}
\item[$\forall p \in P_0.~ p_{1} \:\bot\: p$.] In this case,
we have that $\sigma_{1}(l_{1}) =
(t[x,\ldots,x]_{P_0})|_{p_{1}}$ and, by definition of
descendant, $P_0 = P_1$. Therefore $t[x,\ldots,x]_{P_0}
\rightarrow_{p_{1},R_{1}} t''[x,\ldots,x]_{P_0}$, and
the claim follows by applying the inductive hypothesis to the
sequence
\[ B = (t''[s,\ldots,s]_{P_0} \rightarrow_{p_{2},R_{2}} \cdots
\rightarrow_{p_{n},R_{n}} t'[s,\ldots,s]_{P_n}), \]
where $P_n = P_0 \backslash B$.

\item[$\exists p \in P_0.~p_{1} < p$.]
Since $s$ is operation-rooted and $l_{1}$ is a linear pattern,
then there exists a substitution $\sigma'_{1}$ such that
$\sigma'_{1}(l_{1}) = (t[x,\ldots,x]_{P_0})|_{p_{1}}$ (i.e.,
$\{x \mapsto s \} \circ \sigma'_{1} = \sigma_{1}$).
Therefore, the reduction step
\[ t[x,\ldots,x]_{P_0} \rightarrow_{p_{1},R_{1}} t''[x,\ldots,x]_{P_1} \]
exists and the claim follows by applying the inductive hypothesis
to
\[ B = (t''[s,\ldots,s]_{P_1}
\rightarrow_{p_{2},R_{2}} \cdots \rightarrow_{p_{n},R_{n}}
t'[s,\ldots,s]_{P_n}), \]
where $P_n = P_1 \backslash B$.
\end{description}
\end{description}
\end{proof}

\begin{lemma} \label{liftaux2}
Let $\cR$ be a program.
Let $\theta = \{x_{1} \mapsto s_{1},\ldots,x_{m} \mapsto s_{m} \}$ be
an idempotent substitution such that $s_{i}$ is an operation-rooted
term for all $i = 1,\ldots,m$. Let $s$ be an
operation-rooted term and
$\theta(s) = t_0 \rightarrow_{p_{1},R_{1}} \cdots
\rightarrow_{p_{n},R_{n}} t_n$
be a reduction sequence
where $A_i = (t_{i-1} \rightarrow_{p_{i},R_{i}} t_i)$ and
$P_i = P_{i-1} \backslash A_i$, for $i = 1,\ldots,n$, $n \geq 0$.
If $p \not\leq p_i$ for all $p \in P_{i-1}$, $i = 1,\ldots,n$,
then there exists a reduction sequence
$s \rightarrow_{p_{1},R_{1}} \cdots
\rightarrow_{p_{n},R_{n}} s'$
such that $\theta(s') = t_n$.
\end{lemma}

\begin{proof}
By induction on the number $m$ of bindings in $\theta$:
\begin{description}
\item[\rm Base case.] Consider $\theta = \{ x_{1} \mapsto s_{1} \}$.
We have  $\theta(s) = t_0[s_{1},\ldots,s_{1}]_{P}$ and $s =
t_0[x_{1},\ldots,x_{1}]_{P}$, where $P = \{ p \in \pos(s) \mid s|_{p} =
x_{1} \}$. Then, the claim follows directly by Lemma~\ref{liftaux}.

\item[\rm Induction step.] Consider $\theta = \theta_{1} \cup
\theta'$, where $\theta_{1} = \{ x_{1} \mapsto s_{1} \}$ and
$\theta' = \{x_{2} \mapsto s_{2},\ldots,x_{m} \mapsto s_{m} \}$.
Then, $\theta(s) = t_0[s_{1},\ldots,s_{1}]_{P}$ and $\theta'(s) =
t_0[x_{1},\ldots,x_{1}]_{P}$, where $P = \{ p \in \pos(s)
\mid s|_{p} = x_{1} \}$. Applying Lemma~\ref{liftaux}, we have that
\[
t_0[x_{1},\ldots,x_{1}]_{P} \rightarrow_{p_{1},R_{1}} \cdots
\rightarrow_{p_{n},R_{n}} s''
\] 
is a reduction sequence
such that $\theta_{1}(s'') = t_n$.
By applying the inductive hypothesis to this
derivation, we have that $s \rightarrow_{p_{1},R_{1}} \cdots
\rightarrow_{p_{n},R_{n}} s'$ is a reduction
sequence such that $\theta'(s') = s''$. Therefore, since
$\sdom(\theta_{1}) \cap \sdom(\theta') = \emptyset$, we get
$\theta(s') =
(\theta_{1} \circ \theta')(s') = \theta_{1}(s'') = t_n$,
which proves the claim.
\end{description}
\end{proof}
Now we are ready to extend the  lifting lemma for needed narrowing
(Lemma~\ref{lifting}) to non-constructor substitutions which
do not introduce needed redexes.

\begin{theorem} \label{lifting2}
Let $\cR$ be an inductively sequential program. Let $\sigma$ be a
substitution and $V$ a finite set of variables. Let $s$ be
an operation-rooted term and $\var(s) \subseteq V$. Let
$\sigma(s) \rightarrow_{p_{1},R_{1}} \cdots \rightarrow_{p_{n},R_{n}} t$
be an outermost-needed rewrite sequence
such that, for all root-needed redex $\sigma(s)|_{p}$ of $\sigma(s)$,
$p \in \fpos(s)$.
Then, there exists a needed narrowing derivation
$s \leadsto_{p_{1},R_{1},\sigma_{1}} \cdots
\leadsto_{p_{n},R_{n},\sigma_{n}} t'$ and a substitution $\sigma'$
such that $\sigma'(t') = t$ and
$(\sigma' \circ \sigma_{n} \circ \cdots \circ \sigma_{1} = \sigma)~[V]$.
\end{theorem}

\begin{proof}
We consider two cases:
\begin{description}
\item[\rm $\sigma$ is a constructor substitution.] In this case,
the claim follows directly by applying Lemma~\ref{lifting}, and
$\sigma'$ is a constructor substitution too.

\item[\rm $\sigma$ is a non-constructor substitution.] Then, there
exist substitutions $\theta_{1}$ and $\theta_{2}$ such that $\sigma =
\theta_{2} \circ \theta_{1}$, the substitution $\theta_{1}$ is constructor,
and for all $x \mapsto s' \in \theta_{2}$, $s'$ is operation-rooted.
Then $\sigma(s) = \theta_{2}(\theta_{1}(s))$. By applying
Lemma~\ref{liftaux2}, we have  $\theta_{1}(s)
\rightarrow_{p_{1},R_{1}} \cdots \rightarrow_{p_{n},R_{n}} s''$
such that $\theta_{2}(s'') = t$. On the other hand, since $\sigma$ does not
introduce root-needed redexes (i.e., if $\sigma(s)|_p$ is a
root-needed redex then $p \in \fpos(s)$), then the sequence is
an outermost-needed derivation.
Now, applying Lemma~\ref{lifting} to this reduction sequence, there
exists a needed narrowing derivation $s
\leadsto_{p_{1},R_{1},\sigma_{1}} \cdots
\leadsto_{p_{n},R_{n},\sigma_{n}} t'$ and a
constructor substitution $\sigma''$ such that $\sigma''(t') = s''$
and $(\sigma'' \circ \sigma_{n} \circ \cdots \circ \sigma_{1} =
\theta_{1})~[V]$. By taking $\sigma' = \theta_{2} \circ \sigma''$, we
have $\sigma'(t') = (\theta_{2} \circ \sigma'')(t') =
\theta_{2}(\sigma''(t')) = \theta_{2}(s'') = t$. Finally,
since $\sigma'' \circ \sigma_{n} \circ \cdots \circ \sigma_{1} =
\theta_{1})~[V]$, we have $(\theta_{2} \circ \sigma'' \circ \sigma_{n}
\circ \cdots \circ \sigma_{1} = \theta_{2} \circ \theta_{1}~[V]$, and
hence $(\sigma' \circ \sigma_{n} \circ \cdots \circ \sigma_{1} =
\sigma)~[V]$, which completes the proof.
\end{description}
\end{proof}
The next lemma establishes a strong correspondence between
the closedness of an
expression $t$ and that of one renaming of $t$.

\begin{lemma} \label{indep}
Let $S$ be a finite set of terms, $\rho$ an independent renaming
of $S$, and $S' = \rho(S)$. Given a term $t$,  $ren_{\rho}(t)$ is
$S'$-closed iff $t$ is $S$-closed.
\end{lemma}

\begin{proof}
By induction on the structure of the terms.
\end{proof}

The following lemma states that, if some term $t$ has an
operation-rooted subterm $s$ that contains a redex which is
root-needed
in $t$, then the outermost-needed redex in $s$ is also root-needed
in $t$.

\begin{lemma} \label{easy1}
Let $\cR$ be an inductively sequential program and $t$ be a term. If $s$ is
an operation-rooted subterm of $t$ which
contains a root-needed redex in $t$, then
every outermost-needed redex in $s$ is root-needed in $t$.
\end{lemma}

\begin{proof}
Since $t$ contains at least a root-needed redex, $t$ is not
root-stable. If $t$ has no root-stable form, then every redex in $t$ is
root-needed. Therefore, we assume that $t$ has a root-stable reduct.
If $s$ contains an outermost-needed redex, then, by hypothesis and by Theorem
\ref{TheoNoRootNeededRedex}, $s$ is not root-stable. Hence, by Theorem
\ref{TheoOutNeedIndAreRootNeeded}, such a redex is root-needed in $s$.
By Theorem \ref{TheoRootNeededInSubtermAndRootNeededTerm}, the
conclusion follows.
\end{proof}

The following lemma is helpful.

\begin{lemma} \label{easy2}
Let $\cR$ be an inductively sequential program and $t$ be a term. If $s$ is an
operation-rooted subterm of $t$ which
contains a root-needed redex in $t$ and there is a subterm $s'$ of
$s$ which does not
contain any root-needed redex in $t$, then there is
no outermost-needed derivation from $s$ to a root-stable form which
contracts any redex (or residual) in $s'$.
\end{lemma}

\begin{proof}
If $s$ is root-stable, the claim is trivially true. If $s$ is not
root-stable and there is an outermost-needed derivation starting from $s$
which contracts a (residual of a) redex $s''$ in $s'$, then, by Theorem
\ref{TheoOutNeedIndAreRootNeeded} such a redex is root-needed in $s$.
Therefore, by Theorem
\ref{TheoRootNeededInSubtermAndRootNeededTerm}, $s''$ is root-needed
in $t$, thus leading to a contradiction.
\end{proof}

\begin{proposition} \label{main}
Let $\cR$ be an inductively sequential program. Let $e$ be an equation,
$S$ a finite set of operation-rooted
terms, $\rho$ an independent renaming of $S$,
and $\cR'$ a NN-PE of $\cR$ w.r.t.\ $S$ (under $\rho$)
such that $\cR' \cup \{e'\}$
is $S'$-closed, where $e' = ren_{\rho}(e)$ and $S' = \rho(S)$.
If $e \rightarrow^{\ast} true$ in $\cR$ then
$e' \rightarrow^{\ast} true$ in $\cR'$.
\end{proposition}
\begin{proof}
Since $e'$ is $S'$-closed, by Lemma~\ref{indep}, $e$ is $S$-closed.
Now we prove that, for any reduction sequence $e \rightarrow^{\ast}
true$ in $\cR$ for an $S$-closed term $e$ (not necessarily an
equation), there exists a reduction
sequence $e' \rightarrow^{\ast} true$ in $\cR'$ with $e' =
ren_{\rho}(e)$. Let $B_1,\ldots,B_j$ be all possible needed
reduction sequences from $e$ to $true$ and
$k_i$ the number of contracted redexes
in $B_i$, $i = 1,\ldots,j$. We prove the claim
by induction on the maximum number $n = max(k_1,\ldots,k_j)$
of contracted needed redexes which are necessary to
reduce $e$ to $true$.
\begin{description}
\item[$n = 0$.] This case is trivial since
$e' = ren_{\rho}(true) = true$.

\item[$n > 0$.] Since $e$ is $S$-closed, there exists a closure set
$\{ (p_{1},s_{1}),\ldots,(p_{m},s_{m}) \} \in CSet(S,e)$, $m > 0$,
where $p_{i} \in \pos(e)$ and $s_{i} \in S$, $i=1,\ldots,m$.
Since $e$ contains at least one
needed redex, there exists some $i \in \{1,\ldots,m\}$ such
that $e|_{p_{i}} = \theta(s_{i})$ and
the following facts hold:
\begin{itemize}
\item there exists at least one position $q \in
\pos(e|_{p_{i}})$ such that $e|_{p_{i}.q}$ is a needed redex in $e$,
and
\item for all needed redex $e|_{p_{i}.q'}$ in $e$,
we have $q' \in \fpos(s_{i})$.
\end{itemize}
Informally, $p_i$ addresses an ``innermost'' subterm of $e$
(according to the partition imposed by the closure set) in the
sense that $e|_{p_i}$ contains at least one needed redex and
there is no inner subterm $e|_{p_j}$, $p_i < p_j$, which
contains needed redexes.
Since both $e$ and $e|_{p_{i}}$ are operation-rooted terms,
by Lemma~\ref{easy1} we know that each outermost-needed redex in
$e|_{p_{i}}$
is also a needed redex in $e$ (note that, for derivations
$e\to^\ast true$ in confluent TRSs, the notions of neededness and
root-neededness coincide,
since $true$ is the only root-stable form of $e$).
Let us assume that $q_{1}$ is the position of
such an outermost-needed redex.
Since there is no inner subterm $e|_{p_j}$, $p_i < p_j$, which
contains needed redexes in $e$, then by Lemma~\ref{easy2}, we have
that there is no inner subterm $e|_{p_j}$, $p_i < p_j$, which
contains root-needed redexes in $e|_{p_i}$.
Hence, we can consider a reduction sequence
\[ e[\theta(s_{i})]_{p_{i}} \rightarrow_{p_{i}.q_{1},R_{1}} \ldots
\rightarrow_{p_{i}.q_{k},R_{k}} e[s'_i]_{p_{i}} \rightarrow^\ast true \]
such that the corresponding sequence for $\theta(s_{i})$
\[ \theta(s_{i}) \rightarrow_{q_{1},R_{1}}
\ldots \rightarrow_{q_{k-1},R_{k-1}} s''_i \rightarrow_{q_{k},R_{k}} s'_i \]
is outermost-needed, $s'_i$ is root-stable, and $s''_{i}$ is not
root-stable, $k > 0$.

Now, we prove that $s'_{i}$ is constructor-rooted.
Assume that $s'_{i}$ is operation-rooted. Then, since $t' =
e[s'_{i}]_{p_{i}}$ is root-normalizing, by
Theorem~\ref{TheoRootStabSubtAndRed}, there exists a reduction
sequence $t' \to^{\ast} true$ which does not reduce $s'_{i}$.
Since $s'_{i}$ is operation-rooted and $\cR$ is constructor-based,
then there exists a reduction sequence $e[x]_{p_{i}} \to^{\ast}
true$, with $x \not\in \var(e)$. Therefore, $e \to^{\ast} true$
without reducing $e|_{p_{i}}$, which contradicts the initial
hypothesis that $e|_{p_{i}}$ contains a root-needed redex in $e$.
Hence, $s'_{i}$ is constructor-rooted.

Let $V$ be a finite set of variables containing $\var(s_{i})$.
By Theorem~\ref{lifting2}, we know that
there exists a needed narrowing
derivation $s_i \leadsto_{q_{1},R_{1},\sigma_{1}} \ldots
\leadsto_{q_{k},R_{k},\sigma_{k}} s_i''$ which contracts
the same positions using the same rules and in the same order.
By definition of NN-PE, some resultant of $\cR'$ derives
from a prefix of this needed narrowing derivation.
Assume that the following subderivation
\[ s_i \leadsto_{q_{1},R_{1},\sigma_{1}} \ldots
\leadsto_{q_{j},R_{j},\sigma_{j}} t',~~~ 0 < j \leq k \]
is the one which has been used to construct such a resultant.
Let $\sigma'' = \sigma_j \circ \cdots \circ \sigma_1$. Since
$\theta(s_{i}) \rightarrow_{q_{1},R_{1}}
\ldots \rightarrow_{q_{j},R_{j}} t$,
again by Theorem~\ref{lifting2}, there exists a
substitution $\sigma'$ such that $\sigma'(t') = t$ and
$(\sigma' \circ \sigma'' = \theta) ~[V]$.
Thus, the considered resultant has the form
\[ R' = (\sigma'' (\rho(s_{i})) \rightarrow
ren_{\rho}(t')) \]
and the considered reduction sequence in $\cR$ has the
form
\[ e = e[\theta(s_{i})]_{p_{i}} \rightarrow_{p_{i}.q_{1},R_{1}} \ldots
\rightarrow_{p_{i}.q_{j},R_{j}} e[t]_{p_{i}} \rightarrow^\ast true \]
Now, we prove that $e'$ can be reduced at position $p'_i$
using $R'$, where $p'_i$ is the
corresponding position of $p_{i}$ in $e$ after renaming.
By construction, $\theta(x)$ is $S$-closed for all
$x \in \sdom(\theta)$. Moreover, since $\sigma''$ is constructor
and $(\sigma' \circ \sigma'' = \theta)~[V]$,
we have that $\sigma'(x)$ is also $S$-closed for all
$x \in \sdom(\sigma')$. Then, there exists a substitution
$\theta' = \{ x \mapsto ren_{\rho}(\sigma'(x)) \mid
x \in \sdom(\sigma') \}$ such that $\theta'(x)$ is $S'$-closed
for all $x \in \sdom(\theta')$. By definition of post-processing
renaming, $e'|_{p'_i} = ren_{\rho}(e|_{p_i}) =
ren_{\rho}(\theta(s_i))$. Since $\var(s_i) = \var(\rho(s_i))$ and
$\sigma''$ is constructor, we have $ren_{\rho}(\theta(s_i)) =
ren_{\rho}(\sigma' \circ \sigma''(s_i)) =
\theta'(\sigma''(ren_{\rho}(s_i))) = \theta'(\sigma''(\rho(s_i)))$.
Therefore, the following rewrite step can be proved
\[ e'|_{p'_{i}} = \theta'(\sigma''(\rho(s_{i})))
\rightarrow_{\toppos,R'}
\theta'(ren_{\rho}(t')) = ren_{\rho}(\sigma'(t')) =
ren_{\rho}(t) \]
and thus $e' \rightarrow_{p'_{i},R'}
e'[ren_{\rho}(t)]_{p'_{i}}$. Then, it is immediate to see that
$e'[ren_{\rho}(t)]_{p'_{i}}$ $= ren_{\rho}(e[t]_{p_{i}})$.

Let us now consider the $S$-closedness of $e[t]_{p_{i}}$.
Since $\cR'$ is $S'$-closed, $ren_{\rho}(t')$ is also $S'$-closed.
By Lemma~\ref{indep}, $t'$ is $S$-closed. Since $\sigma'(x)$ is
$S$-closed for all $x \in \sdom(\sigma')$, by definition of
closedness, $\sigma'(t') = t$ is also $S$-closed.
Now we distinguish two cases:
\begin{description}
\item[$p_i = \toppos$.]
Then $e[t]_{p_{i}}$ is trivially $S$-closed since
$t$ is $S$-closed.
\item[$p_i \neq \toppos$.] Let
$j \in \{1,\ldots,m\}$ such that $p_j < p_i$ and there
is no $k \in \{1,\ldots,m\}$ with $p_j < p_k < p_i$.
Let $e|_{p_j} = \gamma(s_j)$ where $y \mapsto s_i \in \gamma$, and
consider the set $P_y = \{ p_{j}.q \in \{
p_{1},\ldots,p_{m} \} \mid s_{j}|_{q} = y \}$. Now we have
two possibilities:
\begin{description}
\item[\rm $P_y$ is a singleton.]
Then $e[t]_{p_{i}}$ is trivially $S$-closed, since
$(p_i,s_{i}) \in CSet(S,e)$ and $t$ is $S$-closed.
\item[\rm $P_y$ is not a singleton.] In this case, we have
$e = e[\theta(s_{i}),\ldots,\theta(s_{i})]_{P_{y}}$.
By considering again the reduction sequences
$\theta(s_{i}) \rightarrow^{\ast} t$ for each $s_{i}$, we get
\[ e[\theta(s_{i}),\ldots,\theta(s_{i})]_{P_{y}}
\rightarrow \cdots \rightarrow e[t,\ldots,t]_{P_{y}} \]
and, by definition of closedness, it is immediate to see that
$e[t,\ldots,t]_{P_{y}}$ is $S$-closed. Moreover, we can construct
the following reduction sequence:
\[
e'[\theta'(\sigma''(\rho(s_{i}))),\ldots,
\theta'(\sigma''(\rho(s_{i})))]_{P'_{y}}
\rightarrow \ldots \rightarrow
e'[ren_{\rho}(t),\ldots,ren_{\rho}(t)]_{P'_{y}}
\]
where $P'_{y}$ corresponds to the positions of $P_{y}$
in $e$ after renaming. Then, we have
\[e'[ren_{\rho}(t),\ldots,ren_{\rho}(t)]_{P'_{y}} =
ren_{\rho}(e[t,\ldots,t]_{P_{y}}).\]
\end{description}
\end{description}
Putting all pieces together, we conclude that there exists
a reduction sequence
\[ e \rightarrow^{+} e[t,\ldots,t]_{P} \rightarrow^{\ast} true \]
in $\cR$, where $P =\{ p_i \}$ or $P = P_y$, such that there exists
a reduction sequence
\[ e' \rightarrow^{+} e'[ren_{\rho}(t),\ldots,ren_{\rho}(t)]_{P'} \]
in $\cR'$, where $P' = \{ p'_i \}$ or $P' = P'_y$, respectively.
Since $e \rightarrow^{+} e[t,\ldots,t]_{P}$ has reduced at least
one needed redex in $e$ and
$e'[ren_{\rho}(t),\ldots,ren_{\rho}(t)]_{P'} =
ren_{\rho}(e[t,\ldots,t]_{P})$, by applying the induction
hypothesis to $e[t,\ldots,t]_{P} \rightarrow^{\ast} true$ in $\cR$,
we get
\[ e'[ren_{\rho}(t),\ldots,ren_{\rho}(t)]_{P'}
\rightarrow^{\ast} true \]
in $\cR'$. By composing  this sequence with the
previous sequence
\[ e' \rightarrow^{+} e'[ren_{\rho}(t),\ldots,ren_{\rho}(t)]_{P'} \]
we get the desired result.
\end{description}
\end{proof}
The completeness of NN-PE is a direct consequence of the
previous proposition
and the soundness and completeness of needed narrowing.

\begin{theorem}[completeness] \label{wcompleteness}
Let $\cR$ be an inductively sequential program. Let $e$ be an equation,
$V \supseteq \var(e)$ a finite set of variables, $S$ a
finite set of operation-rooted terms, and $\rho$ an
independent renaming of $S$. Let $\cR'$ be a
NN-PE of $\cR$ w.r.t.\ $S$ (under $\rho$)
such that $\cR' \cup \{e'\}$ is $S'$-closed,
where $e' = ren_{\rho}(e)$ and $S' = \rho(S)$. If $e
\leadsto_{\sigma}^{\ast} true$ is a needed narrowing
derivation for $e$ in $\cR$, then there exists a needed
narrowing derivation $e' \leadsto_{\sigma'}^{\ast} true$ in $\cR'$ such
that $(\sigma' \leq \sigma)~[V]$.
\end{theorem}

\begin{proof}
Since $e \leadsto_{\sigma}^{\ast} true$, by the soundness of needed
narrowing (claim 1 of Theorem~\ref{theo-needed-properties}),
we have $\sigma(e) \rightarrow^{\ast} true$. Since $e'$ is
$S'$-closed and $\sigma$ is constructor, by definition of closedness,
$\sigma(e')$ is also $S'$-closed and $\sigma(e') = ren_{\rho}(\sigma(e))$.
By Proposition~\ref{main}, there exists a rewrite sequence
$\sigma(e') \rightarrow^{\ast} true$ in $\cR'$. Therefore, since
$\sigma$ is a solution of $e'$ in $\cR'$ and $\cR'$ is inductively
sequential (Theorem~\ref{theo-nn-pe-is-terms}), by the completeness of needed
narrowing (claim 2 of Theorem~\ref{theo-needed-properties}),
there exists a needed narrowing derivation
$e' \leadsto_{\sigma'}^{\ast} true$ such that $(\sigma' \leq \sigma)~[V]$.
\end{proof}

\subsection{Strong Correctness}

Finally, the strong correctness of the transformation
can be easily proved as a direct consequence of
Theorems~\ref{wsoundness} and \ref{wcompleteness}, together
with the independence of solutions computed by
needed narrowing. \\

\noindent
{\it Theorem \ref{theo-nnpe-strong-correct} (strong correctness)}\\
Let $\cR$ be an inductively sequential program. Let $e$ be an equation,
$V \supseteq \var(e)$ a finite set of variables, $S$ a
finite set of operation-rooted terms, and $\rho$ an
independent renaming of $S$. Let $\cR'$ be a
NN-PE of $\cR$ w.r.t.\ $S$ (under $\rho$)
such that $\cR' \cup \{e'\}$ is $S'$-closed,
where $e' = ren_{\rho}(e)$ and $S' = \rho(S)$. Then, $e
\leadsto_{\sigma}^{\ast} true$ is a needed narrowing
derivation for $e$ in $\cR$ iff there exists a needed
narrowing derivation $e' \leadsto_{\sigma'}^{\ast} true$ in $\cR'$ such
that $(\sigma' = \sigma)~[V]$ (up to renaming).

\begin{proof}
We consider the two directions separately:
\begin{description}
\item[\rm Strong soundness.]
We prove the claim by contradiction. Assume that there exists some
substitution $\sigma'$ computed by  needed narrowing for
$e'$ in $\cR'$ such that there is no substitution $\theta$ computed
by  needed narrowing for $e$ in $\cR$ with
$(\theta = \sigma') ~[V]$ (up to renaming).

By Theorem~\ref{wsoundness} (soundness of NN-PE)
and the assumption above, we conclude
that there must be some substitution $\sigma$ computed by
needed narrowing for $e$ in $\cR$ such that
$(\sigma < \sigma') ~[V]$.
Then, by Theorem~\ref{wcompleteness}, there exists
a substitution
$\theta'$ computed by  needed narrowing for $e'$ in $\cR'$
such that $(\theta' \leq \sigma) ~[V]$.
Since $(\theta' \leq \sigma) ~[V]$ and $(\sigma < \sigma') ~[V]$,
we have $(\theta' < \sigma') ~[V]$ which contradicts the independence of
solutions computed by needed narrowing (claim 3 of
Theorem~\ref{theo-needed-properties}).

\item[\rm Strong completeness.]
The proof is perfectly analogous, by considering the completeness of NN-PE
(Theorem~\ref{wcompleteness}) rather than its soundness 
(Theorem~\ref{wsoundness}).
\end{description}
\end{proof}

\end{document}